\pdfoutput=1
\RequirePackage{amsmath,amssymb,mathtools}

\documentclass[twocolumn,epjc3]{svjour3}

\usepackage{graphicx}
\usepackage{lineno}
\usepackage{xcolor}
\usepackage{xspace}
\usepackage[citecolor=blue,linkcolor=blue,linktoc=page,urlcolor=blue,colorlinks=true]{hyperref}
\usepackage{cite}

\newcommand{\nnlojet}{\texorpdfstring{NNLO\protect\scalebox{0.8}{JET}}{NNLOJET}\xspace}

\newcommand{\rd}{\ensuremath{\mathrm{d}}}

\newcommand{\Pp}{{\ensuremath{\mathrm{p}}}\xspace}
\newcommand{\PZ}{{\ensuremath{\mathrm{Z}}}\xspace}
\newcommand{\PW}{{\ensuremath{\mathrm{W}}}\xspace}
\newcommand{\jet}{\text{jet}\xspace}

\newcommand{\mur}{\ensuremath{\mu_{\rm R}}\xspace}
\newcommand{\muf}{\ensuremath{\mu_{\rm F}}\xspace}

\newcommand{\as}{\ensuremath{\alpha_{\mathrm{s}}}\xspace}
\newcommand{\asmz}{\ensuremath{\as(M_{\PZ})}\xspace}
\newcommand{\asmur}{\ensuremath{\as(\mur)}\xspace}
\newcommand{\tilmu}{\ensuremath{\tilde{\mu}}\xspace}

\newcommand{\chisq}{\ensuremath{\chi^{2}}\xspace}
\newcommand{\ndf}{\ensuremath{n_\mathrm{dof}}\xspace}
\newcommand{\ptjet}{\ensuremath{p_\mathrm{T,jet}}\xspace}

\newcommand{\GeVsq}{\ensuremath{\,\mathrm{GeV}^2}\xspace}
\newcommand{\GeV}{\ensuremath{\,\mathrm{GeV}}\xspace}
\newcommand{\pt}{\ensuremath{p_\mathrm{T}}\xspace}
\newcommand{\Qsq}{\ensuremath{Q^{2}}\xspace}

\begin{document}

%-- overwrite "journal name" placeholder to add preprint #'s for arXiv submission
\renewcommand\makeheadbox{%
  \hfill
  CERN-TH-2019-079,
  CFTP/19-020,
  IPPP/19/44,
  MPP-2019-114,
  ZU-TH 29/19
}

\title{%
  Calculations for deep inelastic scattering using \\
  fast interpolation grid techniques at NNLO in QCD \\
  and the extraction of \as from HERA data
}

\author{%
  D.~Britzger\thanksref{mpi}\and
  J.~Currie\thanksref{durham}\and
  A.~Gehrmann-De~Ridder\thanksref{eth,zurich}\and
  T.~Gehrmann\thanksref{zurich}\and
  E.W.N.~Glover\thanksref{durham}\and
  C.~Gwenlan\thanksref{oxford}\and
  A.~Huss\thanksref{cern}\and
  T.~Morgan\thanksref{durham}\and
  J.~Niehues\thanksref{durham}\and
  J.~Pires\thanksref{lisboa1,lisboa2}\and
  K.~Rabbertz\thanksref{karls}\and
  M.R.~Sutton\thanksref{susx}
}

\raggedbottom

\institute{%
  Max-Planck-Institut f\"ur Physik, F\"ohringer Ring 6, 80805 M\"unchen, Germany \label{mpi}\and
  Institute for Particle Physics Phenomenology, Durham University, Durham, DH1 3LE, United Kingdom \label{durham}\and
  Institute for Theoretical Physics, ETH, Wolfgang-Pauli-Strasse 27, CH-8093 Z\"urich, Switzerland \label{eth}\and
  Physik-Institut, Universit\"at Z\"urich, Winterthurerstrasse 190, CH-8057 Z\"urich, Switzerland \label{zurich}\and
  Department of Physics, The University of Oxford, Oxford, OX1 3PU, United Kingdom\label{oxford}\and
  Theoretical Physics Department, CERN, CH-1211 Geneva 23, Switzerland \label{cern}\and
  CFTP, Instituto Superior T\'{e}cnico, Universidade de Lisboa, P-1049-001 Lisboa, Portugal \label{lisboa1}\and
  LIP, Avenida Professor Gama Pinto 2, P-1649-003 Lisboa, Portugal \label{lisboa2}\and
  Institut f\"ur Experimentelle Teilchenphysik (ETP), Karlsruhe
  Institute of Technology (KIT), Wolgang-Gaede-Str.\ 1, D-76131
  Karlsruhe, Germany\label{karls}\and
  Department of Physics and Astronomy, The University of Sussex, Brighton, BN1 9RH, United Kingdom\label{susx}
}

\date{Received: date / Accepted: date}
% The correct dates will be entered by the editor

\maketitle

% \tableofcontents

%\input{tex/erratum}

%!TEX root = ../applfast.tex

% -----------------------------------------------------------------------
\begin{abstract}
  \sloppy
  The extension of interpolation-grid frameworks for perturbative QCD
  calculations at next-to-next-to-leading order (NNLO) is presented
  for deep inelastic scattering (DIS) processes.  A fast and flexible
  evaluation of higher-order predictions for any \emph{a posteriori}
  choice of parton distribution functions (PDFs) or value of the
  strong coupling constant is essential in iterative fitting
  procedures to extract PDFs and Standard Model parameters as well as
  for a detailed study of the scale dependence.  The APPLfast project,
  described here,
  provides a generic interface between the parton-level Monte Carlo
  program \nnlojet and both the APPLgrid and fastNLO libraries for the
  production of interpolation grids at NNLO accuracy. Details of the
  interface for DIS processes are presented together with the required
  interpolation grids at NNLO, which are made available. 
  They cover numerous inclusive jet measurements by the H1
  and ZEUS experiments at HERA\@. An extraction of the strong coupling
  constant is performed as an application of the use of such
  grids and
  a best-fit value of $\asmz = 0.1178\,(15)_\text{exp}\,(21)_\text{th}$
  is obtained using the HERA inclusive jet cross section data.
\end{abstract}
% -----------------------------------------------------------------------

%!TEX root = ../applfast.tex

% -----------------------------------------------------------------------
\section{Introduction}
\label{intro}
% -----------------------------------------------------------------------

%%-- higher-orders and challenges
Modern calculations of higher-order corrections in perturbative QCD for
predictions of cross sections from collider experiments are computationally very
demanding.
In particular, complicated measurement functions and fiducial phase-space definitions associated with differential cross sections prevent an analytic integration over the final-state kinematics, thus calling for numerical approaches.
Next-to-next-to-leading order
computations for differential cross-section predictions, for example, often require
$\mathcal{O}(10^5)$ CPU hours due to the complicated singularity structure of the
real-emission amplitudes and the delicate numerical cancellations they
entail.
Further challenges arise from the requirement of high precision for important benchmark processes.
Common examples are jet production cross sections in both electron--proton
collisions or $\Pp\Pp$ collisions, the Drell--Yan production of $\PZ$ and $\PW$ bosons, and gauge-boson production in association with jets.

%%-- NNLOJET
The \nnlojet program~\cite{Gehrmann:2018szu} is a recent and continuously developing framework
for the calculation of fully differential cross sections for collider experiments.
It includes a large number of processes calculated at NNLO in perturbative QCD,
implemented in a unified and holistic manner.

%%-- intro: the need for efficient methods
For a detailed study of NNLO predictions and
the estimation of theoretical uncertainties, these
calculations must be repeated with different input conditions.
This includes, for example,
using different values for the strong
coupling \asmz,
different parametrisations for the PDFs,
or different choices
for the factorisation or renormalisation scales.
Computationally even more demanding are fits for the determination of the strong coupling
constant and the parton densities in the proton.

In such fits, comparisons must be performed between the data and the NNLO predictions for the many thousands of points that are drawn from the multidimensional parameter space used in the minimisation.
As such, it is computationally prohibitive to run the full calculation at NNLO
for each required input condition encountered in such a fit.
Applications of this nature therefore critically require an efficient approach to perform the convolution of the partonic hard scattering with PDFs, change the value of the strong coupling constant, and vary the scales.

%%-- APPLgrid and fastNLO to the rescue!
The technique of using a grid to store the perturbative
coefficients stripped of the parton luminosity and factors of the strong coupling constant \as, during the full Monte Carlo integration
allows the convolution with arbitrary PDFs to be performed later  with essentially no additional computational cost.
Variation of \asmz, and the renormalisation and factorisation scales is also
possible.
The grid technique, used in Ref.~\cite{Adloff:2000tq}, is implemented independently in the APPLgrid~\cite{Carli:2005,Carli:2010rw}
and fastNLO~\cite{Kluge:2006xs,Britzger:2012bs}
packages.
The technique works by using interpolation functions to
distribute each single weight from the $x$ and $\mu^2$ phase space of the integration, over
a number of discrete \emph{a priori} determined  nodes in that phase space along with
the relevant interpolating function coefficients. Subsequently summing
over those discrete nodes will therefore reproduce the original value
for the weight, or any product of the weight with some function of the
phase space parameters for that specific phase space point.
One dimension in the grid is required for each parameter upon which the subsequently varied
parameters will depend. For instance, for DIS processes, a dimension
for $x$ and $\mu^2$ will be required. For $\Pp\Pp$ collisions, a third
dimension must be added to account for the momentum fraction $x_2$
of the second proton.

%%-- paper outline
% APPLfast
\sloppy
This paper describes developments in the APPLfast project which provides
a common interface for the APPLgrid and fastNLO grid libraries to link to the \nnlojet program for the calculation of the perturbative coefficients.
% application alpha_s
The generation and application of interpolation grids for
DIS jet production at NNLO~\cite{Currie:2016ytq,Currie:2017tpe} is discussed.
Grids are made publicly available on the \texttt{ploughshare} website~\cite{DIS:grids}.
A subset of these grids have previously been employed for a determination of the
strong coupling constant, \asmz~\cite{Andreev:2017vxu}.
Here, additional details of the grid methodology for DIS are discussed,
together with the NNLO extraction of \asmz using data on inclusive jet
production from both H1 and ZEUS.

%!TEX root = ../applfast.tex

% -----------------------------------------------------------------------
\section{DIS at NNLO and the \nnlojet framework}
\label{nnlojet}
% -----------------------------------------------------------------------

Jet production in the neutral-current DIS process proceeds through the scattering of a parton from the proton with a virtual photon or Z boson that mediates the interaction.
The cross section for this process is given by the convolution of the parton distribution function with the partonic hard-scattering cross section
\begin{align}
  \sigma &=
  \int \rd x \; f_a(x,\muf) \; \rd\hat{\sigma}_a(x,\mur,\muf) \,,
  \label{eq:sigma}
\end{align}
which includes an implicit summation over the index $a$ which denotes the incoming parton flavour.
In perturbative QCD, the hard-scattering cross section can be expanded in the coupling constant
\begin{align*}
  \rd\hat{\sigma}_a(x,\mur,\muf)
  &= \sum_{p} \left(\frac{\as(\mur)}{2\pi}\right)^{k+p}
  \rd\hat{\sigma}^{(p)}_a(x,\mur,\muf)
  \, ,
\end{align*}
where $k$ corresponds to the power in $\as$ at leading order (LO).
Jet cross section measurements in DIS commonly employ a reconstruction in the Breit frame of reference, in which the proton and the gauge boson of virtuality \Qsq collide head-on.
This is further assumed in the remainder of this work.
As a consequence, jet production proceeds through the basic scattering processes $\gamma^*g\to q\bar{q}$ and $\gamma^*q\to qg$, thus requiring at least two partons in the final state. 
This choice not only gives a direct sensitivity to \as ($k=1$) but also a rare handle on the gluon density already at LO\@.

Calculations at higher orders in perturbation theory comprise distinct parton-level ingredients that may involve additional loop integrations and real emission.
For jet production in DIS at NNLO ($p=2$), three types of contributions enter the calculation: 
The double-real (RR) contribution comprising tree-level amplitudes with two additional partons in the final state~\cite{Hagiwara:1988pp,Berends:1988yn,Falck:1989uz},
the real--virtual (RV) contribution that requires one-loop amplitudes with one additional emission~\cite{Glover:1996eh,Bern:1996ka,Campbell:1997tv,Bern:1997sc},
and the double-virtual (VV) contribution involving two-loop amplitudes~\cite{Garland:2001tf,Garland:2002ak,Gehrmann:2011ab}.
Each of these ingredients are separately infrared divergent and only finite after taking their sum, as dictated by the Kinoshita--Lee--Nauenberg theorem.
The different manifestations of the singularities among the three contributions,
related to the distinct parton multiplicities, makes the cancellation of infrared singularities a highly non-trivial task.
Fully differential predictions in particular, require a procedure to re-distribute and cancel the singularities while retaining the information on the final-state kinematics.
The antenna subtraction formalism~\cite{GehrmannDeRidder:2005cm,Daleo:2006xa,Currie:2013vh} accomplishes this by introducing local counter terms with the aim to render each contribution manifestly finite and thus amenable to numerical Monte Carlo integration methods.
The partonic hard-scattering cross section can be schematically written as
\begin{align}
  \int\rd \hat{\sigma}^{(2)}_a
  &= \phantom{+}
  \int_{\Phi^{(n+2)}} \Bigl(
    \rd\hat{\sigma}^{RR}_a - \rd\hat{\sigma}^{S}_a
  \Bigr)
  \nonumber\\&\quad
  +\int_{\Phi^{(n+1)}} \Bigl(
    \rd\hat{\sigma}^{RV}_a - \rd\hat{\sigma}^{T}_a
  \Bigr)
  \nonumber\\&\quad
  +\int_{\Phi^{(n)\phantom{+0}}} \Bigl(
    \rd\hat{\sigma}^{VV}_a - \rd\hat{\sigma}^{U}_a
  \Bigr)
  \, ,
  \label{eq:sigmahat_nnlo}
\end{align}
where the subtraction terms $\rd\hat{\sigma}^{S,T,U}_a$
absorb in their definition the NNLO mass-factorisation terms from the PDFs
and are explicitly given in Ref.~\cite{Currie:2017tpe}.
Note that differential distributions can be accommodated in Eq.~\eqref{eq:sigma} via event selection cuts in the measurement functions that are implicitly contained in $\rd\hat{\sigma}^X_a$.

The \nnlojet framework~\cite{Gehrmann:2018szu} provides the necessary infrastructure to perform calculations at NNLO using the antenna subtraction method following the master formula~\eqref{eq:sigmahat_nnlo} and incorporates all available processes under a common code base.
The parton-level Monte Carlo generator evaluates the integral for each perturbative order ($p=0,\ldots$) by summing over samples of the phase space $(x_m,\Phi_m)_{m=1,\ldots,M_p}$ with their associated weights $w^{(p)}_{a;m}$.
The cross section in Eq.~\eqref{eq:sigma} can then be computed via
\begin{align}
  \sigma \xrightarrow{\text{MC}}
  \sum_{p} \sum_{m=1}^{M_p} &
  \;
  \left(\frac{\as(\mur{}_{;m})}{2\pi}\right)^{k+p}
  \nonumber\\&\times
  f_{a}(x_m, \muf{}_{;m}) \;
  w^{(p)}_{a;m} \; \rd\hat{\sigma}^{(p)}_{a;m} \, ,
  \label{eq:sigma_MC}
\end{align}
using the short-hand notation
\begin{align*}
  \mu_{X;m} &\equiv \mu_X(\Phi_m) \quad \text{for $X=\mathrm{R},\,\mathrm{F}$,}
  \\
  \rd\hat{\sigma}^{(p)}_{a;m} &\equiv
  \rd\hat{\sigma}^{(p)}_{a}(x_m,\mur{}_{;m},\muf{}_{;m}) \,.
\end{align*}
For the interface of the \nnlojet code to the grid-filling tools described in Sect.~\ref{gridtechnique}, additional hook functions are provided that, e.g., allow for a full decomposition of the differential cross section $\rd\hat{\sigma}^{(p)}_{a}$ into the coefficients of the logarithms in the renormalisation and factorisation scales:
\begin{align}
  & \rd\hat{\sigma}^{(p)}_{a} (\mur^2,\muf^2)
  = \sum_{\mathclap{\substack{\alpha, \beta \\ \alpha+\beta\leq p}}}
  \rd\hat{\sigma}^{(p | \alpha,\beta)}_{a}
  \ln^{\alpha}\left(\frac{\mur^2}{\mu^2}\right)
  \ln^{\beta} \left(\frac{\muf^2}{\mu^2}\right)
  \label{eq:weight_decomp}
  \, ,
\end{align}
where $\mu$ is the reference scale of the decomposition.
This ensures maximal flexibility for the interface to accommodate different prescriptions, such as the different strategies pursued by APPLgrid and fastNLO for the reconstruction of the scale dependence.

%!TEX root = ../applfast.tex

% -----------------------------------------------------------------------
\section{The APPLgrid and fastNLO packages}
\label{gridtechnique}
% -----------------------------------------------------------------------

The grid technique allows an accurate approximation of a continuous function $f(x)$ to be 
obtained from the knowledge of its value at discrete nodes 
$a\equiv x^{[0]} < x^{[1]} < \ldots < x^{[N]}\equiv b$ that partition
the interval $[x_{\rm min},x_{\rm max}]$ into $N$ disjoint sub-intervals.
To this end, interpolation kernels $E_i(x)$ are introduced for each node $i$, 
which are constructed from polynomials of degree $n$ and satisfy $E_i(x^{[j]})=\delta_i^j$.
The set of interpolation kernels further form a partition of unity,
\begin{align}
  1 &= \sum_{i=0}^{N} E_i(x) \quad \text{for $a \leq x \leq b$} \,.
\end{align}
As a result, the continuous function $f(x)$ can be approximated as
\begin{align}
  f(x) &\simeq \sum_{i=0}^{N} f^{[i]} \; E_i(x) \quad \text{with $f^{[i]} \equiv f(x^{[i]})$} .
\end{align}

In practice, the interpolation is often set up using equidistant nodes ($x^{[k]}=x^{[0]}+k\,\delta x$)
for simplicity. This can however result into a sub-optimal placement of grid nodes resulting in a poor
interpolation quality, which in turn would require an increase in the number of nodes to
achieve the required target accuracy.
Alternatively, the accuracy can be greatly improved by performing a variable transformation $x \longmapsto  y(x)$ that increases the density of nodes in regions where $f(x)$ varies more rapidly.
In this case, nodes are chosen with respect to $y(x)$ and the corresponding interpolation kernels are denoted by $E^y_i(x)$.

Finally, when the function $f(x)$ appears under an integral, the integration can be approximated by a sum over the nodes $i$,
\begin{align}
  &\int_a^b \rd x \; f(x) \; g(x) %&\simeq
  \simeq
  \sum_{i=0}^{N} f^{[i]} \; g_{[i]}
  \, ,
  \label{eq:conv}
\end{align}
using the definition
\begin{align}
  g_{[i]} &\equiv
  \int_a^b \rd x \;  E_i(x) \; g(x)
  \, .
\end{align}
The time-consuming computation of the integral can then be performed once and for all to
produce a grid $g_{[i]}$ ($i=0,\ldots,N$) and the integral in Eq.~\eqref{eq:conv} can
be approximated for different functions $f(x)$ using the sum from the right hand side,
which can be evaluated very quickly.

\subsection{Application to the DIS cross section}

For DIS processes, the different parton densities $f_a(x,\muf)$ can be included using the grid technique.
In this case, a two-dimensional grid in the two independent variables $x$ and $\muf$ is constructed.
The respective interpolation kernels $E^y_i(x)$ and $E^\tau_j(\muf)$ can be chosen independently for the two variables,
introducing the additional transformation in the scale variable,  $\muf \longmapsto  \tau(\muf)$.
Typical transformations for DIS  are for instance
\begin{align}
  y(x) &= \ln \frac{1}{x} + \alpha (1 - x ) &\text{or}&&
  y(x) &= \ln^\alpha\frac{1}{x}
  \label{eq:yx}
\end{align}
for the momentum fraction, and
\begin{align}
  \tau(\mu) &= \ln \ln \frac{\mu^2}{\Lambda^2} &\text{or}&&
  \tau(\mu) &= \ln\ln \frac{\mu}{\Lambda},
  \label{eq:taumu}
\end{align}
for the hard scale,
where the parameter $\alpha$ can be used to increase the density of nodes at high or low values of $x$ or $\mu$, and $\Lambda$ can be chosen of
the order of $\Lambda_{\mathrm{QCD}}$, but need not necessarily be identical.
Additional transforms are available in both APPLgrid and fastNLO.

For any value of $x$ and $\mu$, both the PDFs and the running of the strong coupling can then be represented by a sum over the interpolation nodes,%
\begin{align}
  \as(\mu) \; f_a(x,\mu) &\simeq
  \sum_{i,j} \as^{[j]} \; f^{[i,j]}_{a} \; E^y_i(x) \; E^\tau_j(\mu)
  \, ,
\end{align}
where $\mur=\muf\equiv\mu$ has been set for simplicity.
The computationally expensive convolution with
the PDFs from Eq.~\eqref{eq:sigma}, which further includes an implicit phase-space dependence through the scale $\mu$, can thus be
approximated by a two-fold summation,
\begin{align}
  \sigma &=
  \sum_{p} \int \rd x \left(\frac{\as(\mu)}{2\pi}\right)^{k+p} f_a(x,\mu) \; \rd\hat{\sigma}^{(p)}_a(x,\mu)
  \nonumber\\&\simeq
  \sum_{p} \sum_{i,j} \biggl(\frac{\as^{[j]}}{2\pi}\biggr)^{k+p} f^{[i,j]}_{a} \; \hat{\sigma}^{(p)}_{a[i,j]}
  \, .
  \label{eq:sigma_grid}
\end{align}
Here, the grid of the hard coefficient function at the perturbative order $p$ has been defined as
\begin{align}
  \hat{\sigma}^{(p)}_{a[i,j]}
  &=
  \int \rd x
  \; E^y_i(x)
  \; E^\tau_j(\mu)
  \; \rd\hat{\sigma}^{(p)}_a(x,\mu)
  \, ,
\end{align}
which can be readily obtained during the Monte Carlo integration as described in Eq.~\eqref{eq:sigma_MC} by accumulating the weights
\begin{align}
  \hat{\sigma}^{(p)}_{a[i,j]} & \xrightarrow{\text{MC}}
  \sum_{m=1}^{M_p}
  E^y_i(x_m)
  \; E^\tau_j(\mu_{m})
  \; w^{(p)}_{a;m}
  \; \rd\hat{\sigma}^{(p)}_{a;m}
  \label{eq:grid_gen}
\end{align}
during the computation.

\subsection{Renormalisation and factorisation scale dependence}

With the hard coefficients $\hat{\sigma}^{(p)}_{a[i,j]}$ determined separately order by order in
$\as$, it is straightforward to restore the dependence on the renormalisation scale, $\mur$,  and
factorisation scale, $\muf$, using the RGE running of $\as$ and the DGLAP evolution for the PDFs.
To this end, any functional form can be chosen that depends on the scale $\mu$ that was used
during the grid generation~\eqref{eq:grid_gen};%
\begin{align}
  \mu_X &= \mu_X(\mu) \quad \text{for $X=\mathrm{R},\,\mathrm{F}$} .
\end{align}
Generating larger grids that include additional alternative central scale choices each with an additional dimension in the grid allows for the scale choice used in the convolution to be any arbitrary function of these independent central scales, $\mu_X = \mu_X(\mathcal{O}_1, \mathcal{O}_2, \ldots)$.
The functionality for storing an additional central scale is implemented in fastNLO but entails an increase in the grid size and therefore also on the memory footprint during the computation.
Using the short-hand notation
\begin{align*}
  L^{[j]}_{\mathrm{X}} &\equiv \ln\left(\frac{\mu_X^2(\mu^{[j]})}{\mu^{2[j]}}\right) \quad \text{for $X=\mathrm{R},\,\mathrm{F}$},
  \nonumber\\
  \as^{[j_{\to\mathrm{R}}]} &\equiv \as(\mur(\mu^{[j]})), { \text{ and\  }}
  f^{[i,j_{\to\mathrm{F}}]}_{a}  \equiv f_a(x^{[i]},\muf(\mu^{[j]})) ,
\end{align*}
the full scale dependence up to  NNLO is given by
\begingroup
\allowdisplaybreaks
\begin{align}
  &\sigma^\text{NNLO}(\mur,\muf) =
  \sum_{i,j} \biggl(\frac{\as^{[j_{\to\mathrm{R}}]}}{2\pi}\biggr)^{k}
  f^{[i,j_{\to\mathrm{F}}]}_{a} \; \hat{\sigma}^{(0)}_{a[i,j]}
  \nonumber\\&\quad
  +\sum_{i,j} \biggl(\frac{\as^{[j_{\to\mathrm{R}}]}}{2\pi}\biggr)^{k+1}
  \biggl\{
    f^{[i,j_{\to\mathrm{F}}]}_{a} \; \hat{\sigma}^{(1)}_{a[i,j]}
    \nonumber\\*&\qquad
    + \Bigl[
      k \beta_0 f^{[i,j_{\to\mathrm{F}}]}_{a} L^{[j]}_{\mathrm{R}}
      \nonumber\\*&\qquad\quad
      -(P^{(0)}\otimes f)^{[i,j_{\to\mathrm{F}}]}_{a} L^{[j]}_{\mathrm{F}}
    \Bigr] \; \hat{\sigma}^{(0)}_{a[i,j]}
  \biggr\}
  \nonumber\\&\quad
  +\sum_{i,j} \biggl(\frac{\as^{[j_{\to\mathrm{R}}]}}{2\pi}\biggr)^{k+2}
  \biggl\{
    f^{[i,j_{\to\mathrm{F}}]}_{a} \; \hat{\sigma}^{(2)}_{a[i,j]}
    \nonumber\\&\qquad
    + \Bigl[
      (k+1) \beta_0 f^{[i,j_{\to\mathrm{F}}]}_{a} L^{[j]}_{\mathrm{R}}
      \nonumber\\*&\qquad\quad
      -(P^{(0)}\otimes f)^{[i,j_{\to\mathrm{F}}]}_{a} L^{[j]}_{\mathrm{F}}
    \Bigr] \; \hat{\sigma}^{(1)}_{a[i,j]}
    \nonumber\\&\qquad
    + \Bigl[
      \Bigl(
        k \beta_1 + \tfrac{1}{2}k(k+1)\beta_0^2 L^{[j]}_{\mathrm{R}}
      \Bigr) \; f^{[i,j_{\to\mathrm{F}}]}_{a} L^{[j]}_{\mathrm{R}}
      \nonumber\\*&\qquad\quad
      -(P^{(1)}\otimes f)^{[i,j_{\to\mathrm{F}}]}_{a} L^{[j]}_{\mathrm{F}}
      \nonumber\\*&\qquad\quad
      +\tfrac{1}{2}(P^{(0)}\otimes P^{(0)}\otimes f)^{[i,j_{\to\mathrm{F}}]}_{a} L^{2[j]}_{\mathrm{F}}
      \nonumber\\*&\qquad\quad
      + \Bigl(
        \tfrac{1}{2} \beta_0 L^{[j]}_{\mathrm{F}}  -(k+1) \beta_0 L^{[j]}_{\mathrm{R}}
      \Bigr)
      \nonumber\\*&\qquad\quad\quad
      \times (P^{(0)}\otimes f)^{[i,j_{\to\mathrm{F}}]}_{a} L^{[j]}_{\mathrm{F}}
    \Bigr] \; \hat{\sigma}^{(0)}_{a[i,j]}
  \biggr\}
  \, .
  \label{eq:scale_ana}
\end{align}
\endgroup
In APPLgrid, this summation
is performed on the fly only if and when required, with the convolutions with the splitting functions $P^{(n)}$
performed using Hoppet~\cite{Salam:2008qg}.

As an alternative to the analytical reconstruction of the scales in Eq.~\eqref{eq:scale_ana},
individual grids for the additional independent coefficients of the scale logarithms can be generated.
This corresponds to the default strategy in the fastNLO library and the full scale dependence can be reconstructed through
\begin{align}
  &\sigma^\text{NNLO}(\mur,\muf) =
  \sum_{i,j} \biggl(\frac{\as^{[j_{\to\mathrm{R}}]}}{2\pi}\biggr)^{k}
  f^{[i,j_{\to\mathrm{F}}]}_{a} \; \hat{\sigma}^{(0|0,0)}_{a[i,j]}
  \nonumber\\&\quad
  +\sum_{i,j} \biggl(\frac{\as^{[j_{\to\mathrm{R}}]}}{2\pi}\biggr)^{k+1}
  f^{[i,j_{\to\mathrm{F}}]}_{a} \;
  \nonumber\\&\quad \quad\times
  \biggl\{
    \hat{\sigma}^{(1|0,0)}_{a[i,j]}
    + L^{[j]}_{\mathrm{R}} \; \hat{\sigma}^{(1|1,0)}_{a[i,j]}
    + L^{[j]}_{\mathrm{F}} \; \hat{\sigma}^{(1|0,1)}_{a[i,j]}
  \biggr\}
  \nonumber\\&\quad
  +\sum_{i,j} \biggl(\frac{\as^{[j_{\to\mathrm{R}}]}}{2\pi}\biggr)^{k+2}
  f^{[i,j_{\to\mathrm{F}}]}_{a} \;
  \nonumber\\&\quad \quad\times
  \biggl\{
    \hat{\sigma}^{(2|0,0)}_{a[i,j]}
    + L^{[j]}_{\mathrm{R}} \; \hat{\sigma}^{(2|1,0)}_{a[i,j]}
    + L^{[j]}_{\mathrm{F}} \; \hat{\sigma}^{(2|0,1)}_{a[i,j]}
    \nonumber\\&\qquad\quad
    + L^{2[j]}_{\mathrm{R}} \; \hat{\sigma}^{(2|2,0)}_{a[i,j]}
    + L^{2[j]}_{\mathrm{F}} \; \hat{\sigma}^{(2|0,2)}_{a[i,j]}
    \nonumber\\&\qquad\quad
    + L^{[j]}_{\mathrm{R}} \; L^{[j]}_{\mathrm{F}} \; \hat{\sigma}^{(2|1,1)}_{a[i,j]}
  \biggr\}
  \, ,
  \label{eq:scale_log}
\end{align}
where the grids are produced in analogy with Eq.~\eqref{eq:grid_gen} but using the decomposition of Eq.~\eqref{eq:weight_decomp}
\begin{align*}
  \hat{\sigma}^{(p|\alpha,\beta)}_{a[i,j]} & \xrightarrow{\text{MC}}
  \sum_{m=1}^{M_p}
  E^y_i(x_m) \; E^\tau_j(\mu_{m})
  \;  w^{(p)}_{a;m}
  \; \rd\hat{\sigma}^{(p|\alpha,\beta)}_{a;m}
  \, .
  \label{eq:grid_gen_decomp}
\end{align*}
Using additional coefficient grids reduces the numerical complexity of
the {\em a posteriori} convolutions involving the splitting functions
and is faster for these terms but increases the number of summations over
the grids for the full NNLO calculation from three to ten. The evaluation of these
additional terms can be performed using the full expressions or they can
be obtained numerically by evaluating the Monte Carlo weights
for six independent scale pairs $(\mur,\muf)$ and solving a linear equation for the coefficients.

%!TEX root = ../applfast.tex

% -----------------------------------------------------------------------
\section{The APPLfast project}
\label{applfast}
% -----------------------------------------------------------------------

The APPLfast project provides a library of code written in C++ with Fortran callable components.
It is a lightweight interface used to bridge between the \nnlojet code and the specific code for
booking and filling the grids themselves using  either APPLgrid or fastNLO.

The basic structure for the filling of either grid technology is essentially the same, and as such,
much of the functionality for the interface exists as common code that is used for
filling both, with only the code that actually fills the weights needing to be  specific
to either technology. Efforts are under way to implement a common filling API for both
fastNLO and APPLgrid, which will allow significantly more of the specific
filling code to be shared.

A design principle, applied from the outset, was that the interface should be as unobtrusive
as possible in the \nnlojet code, and should
provide no additional performance overhead in terms of execution time when not filling
a grid. When filling a grid, any additional overhead should be kept as low as possible.
This is achieved by the use
of a minimal set of hook functions that can be called from within the \nnlojet code itself
and which can be left within the code with no impact on performance if the grid filling
functionality is not required.
The original proof-of-concept implementation accessed
the required variables for the weights, scales and momentum fractions via the \nnlojet
data structures directly, but following this it was decided to instead implement custom access functions
that allow, e.g., for a full decomposition of the event weights as described by Eq.~\eqref{eq:weight_decomp}, thus
enabling a more straightforward design for the filling code.

Each process in \nnlojet consists of a large number of subprocesses.
In order to fill the grids, during the configuration stage
the internal list of \nnlojet processes is mapped to a minimal set of the unique parton
luminosities that are used for the grid.
When filling, these internal \nnlojet %\verb|process-id|
process identifiers are used to determine which
parton luminosity terms in the grid should be filled on the interface side.

Generating a cross section grid using \nnlojet typically involves four stages:
\begin{enumerate}
  \item \emph{Vegas adaption}: This is the first stage in the standard \nnlojet
  workflow and is used to generate an optimised Vegas phase-space grid for the subsequent production
  runs. At this stage the grid filling is not enabled and \nnlojet can run in multi-threaded mode.
  \item \emph{Grid warm-up}: This is required in order
  to optimise the limits for the phase space in $x$ and $\muf$ for the grids.
  During this stage, the \nnlojet code runs in a custom mode
  intended solely to sample the phase-space volume, thus skipping
  the costly evaluation of the Matrix Elements.
  \item \emph{Grid production}: Here, the grids from stage~2 are filled with the
  weights generated from a full \nnlojet run, using the optimised
  phase-space sampling determined in stage~1. The calculation can be
  run in parallel using many independent jobs to achieve the desired
  statistical precision.
  \item \emph{Grid combination}: In this stage, the grids from the individual jobs are combined, first merging the results for each of the LO, NLO (R and V), and NNLO (RR, VV, RV) terms separately, and subsequently assembling the respective grids into a final master grid.
\end{enumerate}
The procedure to combine the interpolation grids closely follows the
one developed for \nnlojet~\cite{Ridder:2016rzm}.  Each cross-section
bin in the observable of each calculated grid is weighted with the
same number as determined by the \nnlojet merging script for the
combination of the final cross sections.

The stabilisation of higher-order cross sections with respect to
statistical fluctuations demands a substantial number of events to be
generated.  This is particularly true for the double-real
contribution, since the large number of final-state partons lead to a
complex pattern of infrared divergences that need to be compensated.
Typically, computing times of the order of hundreds of thousands of
CPU hours are required. In stage~3 it is therefore mandatory to run
hundreds to thousands of separate jobs in parallel, in particular for
the NNLO sub-contributions. The resulting interpolation grids for each
cross section and job typically are about 10--100 MBytes in size.  The
final master grid obtained by summing the output from all jobs then is
somewhat larger than the largest single grid, because it contains at least
one weight grid for each order in $\alpha_s$.

\newcommand{\xnodesfig}{
\newcommand{\tinyvspace}{\vspace{1.35mm}}
\newcommand{\smallhspace}{\hspace{7.4cm}}

\begin{figure}[th]
  \centering
  \begin{minipage}[t]{\textwidth}
  \includegraphics[width=0.5\textwidth]{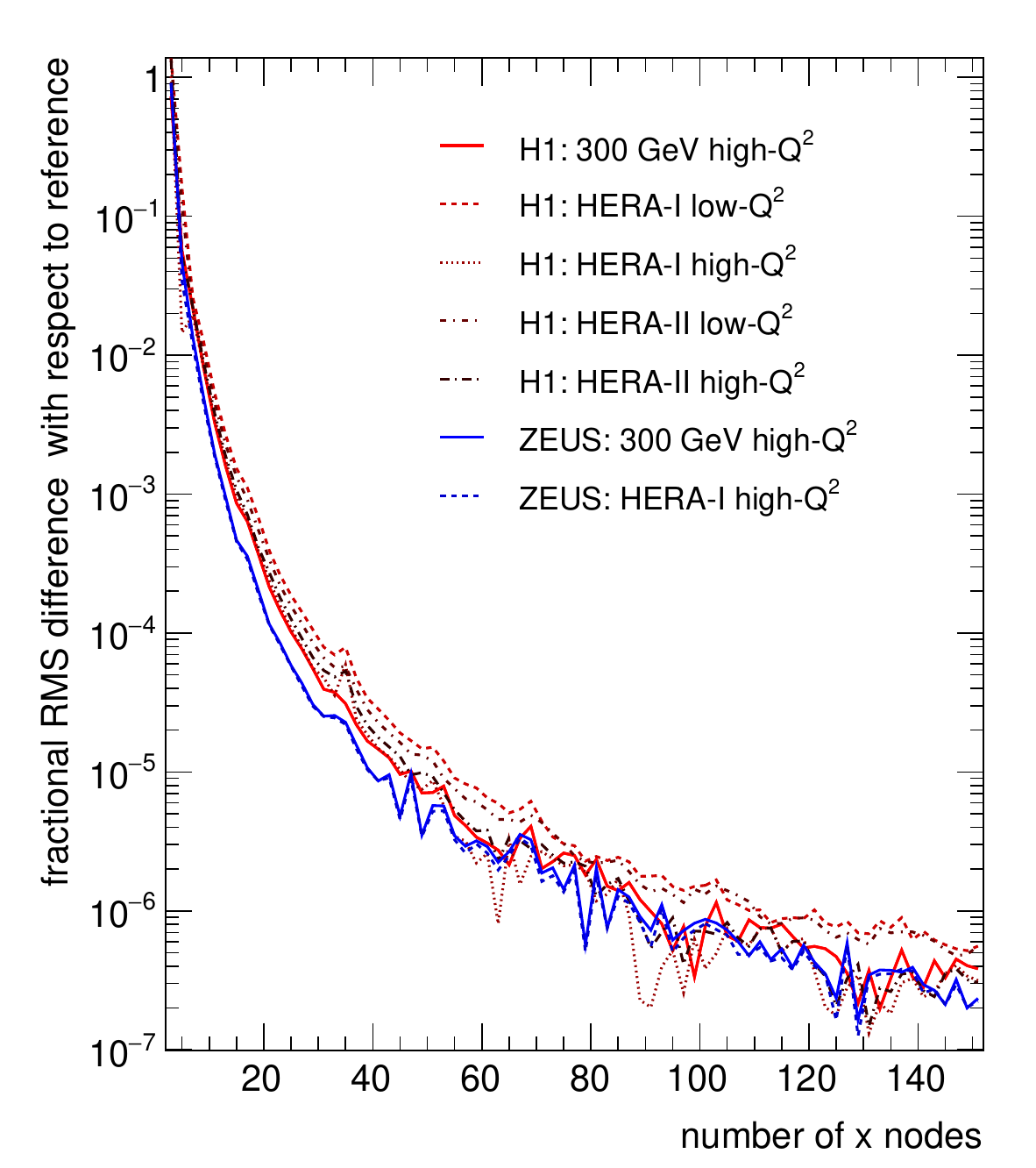}

  \vspace{-88mm}
  \hspace{7.0cm}\cite{Adloff:2000tq}

  \tinyvspace
  \hspace{6.8cm}\cite{Aaron:2010ac}

  \tinyvspace
  \hspace{6.9cm}\cite{Aktas:2007aa}

  \tinyvspace
  \hspace{6.9cm}\cite{Andreev:2016tgi}

  \tinyvspace
  \hspace{7.0cm}\cite{Andreev:2014wwa}

  \tinyvspace
  \hspace{7.3cm}\cite{Chekanov:2002be}

  \tinyvspace
  \hspace{7.2cm}\cite{Chekanov:2006xr}

  \vspace{2cm}
  \end{minipage}
  \caption{%
    The RMS difference between the fast grid convolution and reference histogram as a
    function of the number of grid nodes in momentum fraction, $x$ for the HERA inclusive jet measurements in DIS.
    \label{fig:xnodes}
  }
\end{figure}
}
\xnodesfig

The interpolation accuracy must be evaluated to ensure
that the results of the full calculation can be reproduced with the
desired precision. For sufficiently well-behaved functions, as usually
the case for PDFs, it is always possible to reach such
precision by increasing the number of nodes
in the fractional momentum $x$ and scale $\mu$ at the cost of larger grid sizes.
For proton-proton scattering, because of the additional momentum
fraction associated with the second proton, the grid size grows
quadratically with the number of $x$ nodes.

To optimise the number of nodes necessary to achieve a sufficient
approximation accuracy, several parameters and techniques can be
adapted: Notably, the order or method of interpolation,
the transform used for $x$ and $\mu$, and the accessed ranges in $x$
and $\mu$, as determined in the grid warm-up stage~3, can be chosen
such that the number of nodes can be reduced significantly while
retaining the same approximation accuracy.  Figure~\ref{fig:xnodes}
shows the root mean square (RMS) of the fractional difference of the fast grid
convolution with respect to the corresponding reference for HERA
inclusive jet production data. This uses a third order interpolation
in the transformed $y(x)$ variable and the transform from
Eq.~\eqref{eq:taumu} and shows that the precision is better than one per
mille for grids with 20 $x$ nodes, and better than 0.1 per mille for
grids with more than 30 $x$ nodes.

\newcommand{\closefig}{
\begin{figure*}[tbh]
  \centering
  \includegraphics[width=0.31\textwidth]{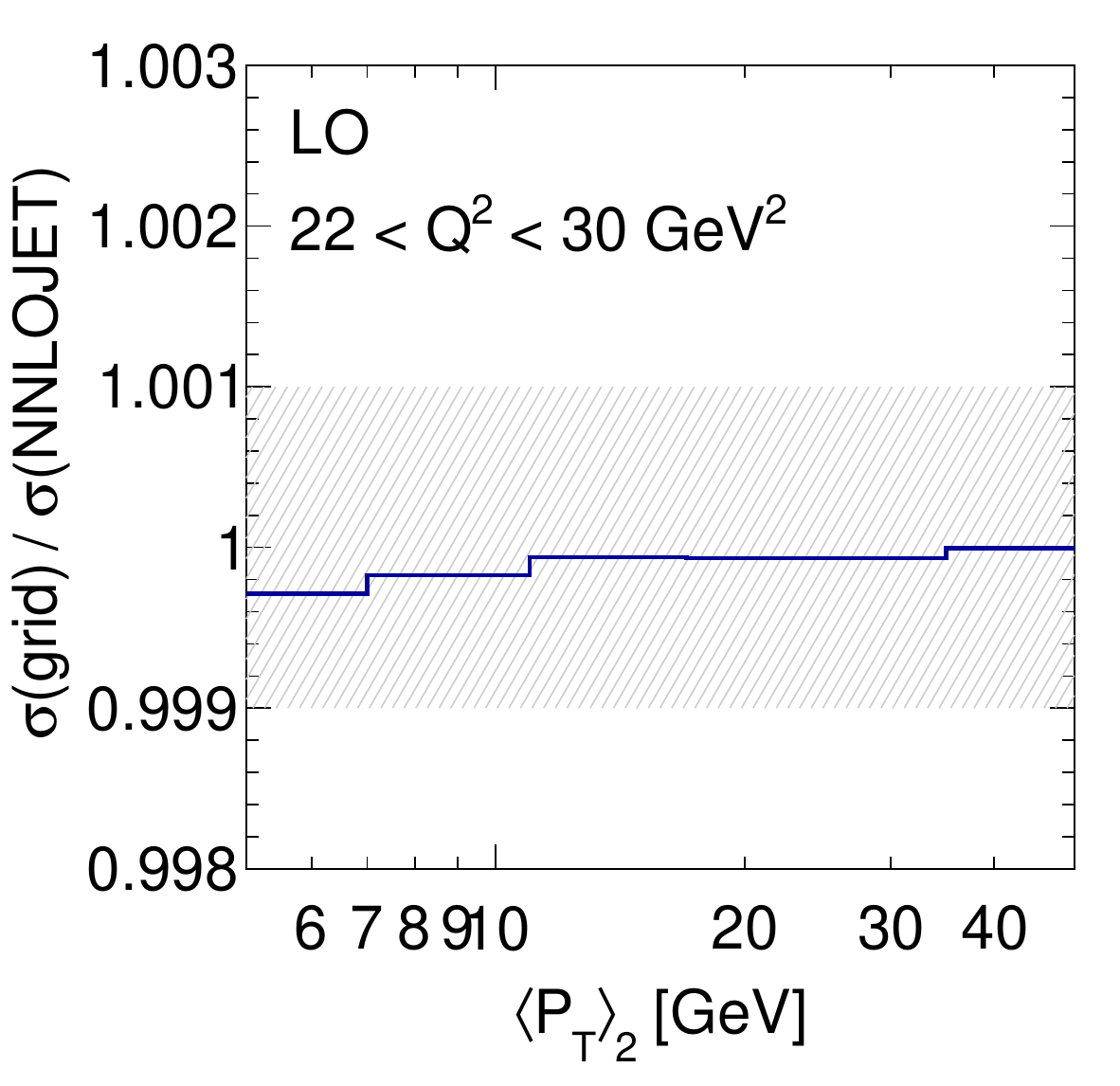}
  \includegraphics[width=0.31\textwidth]{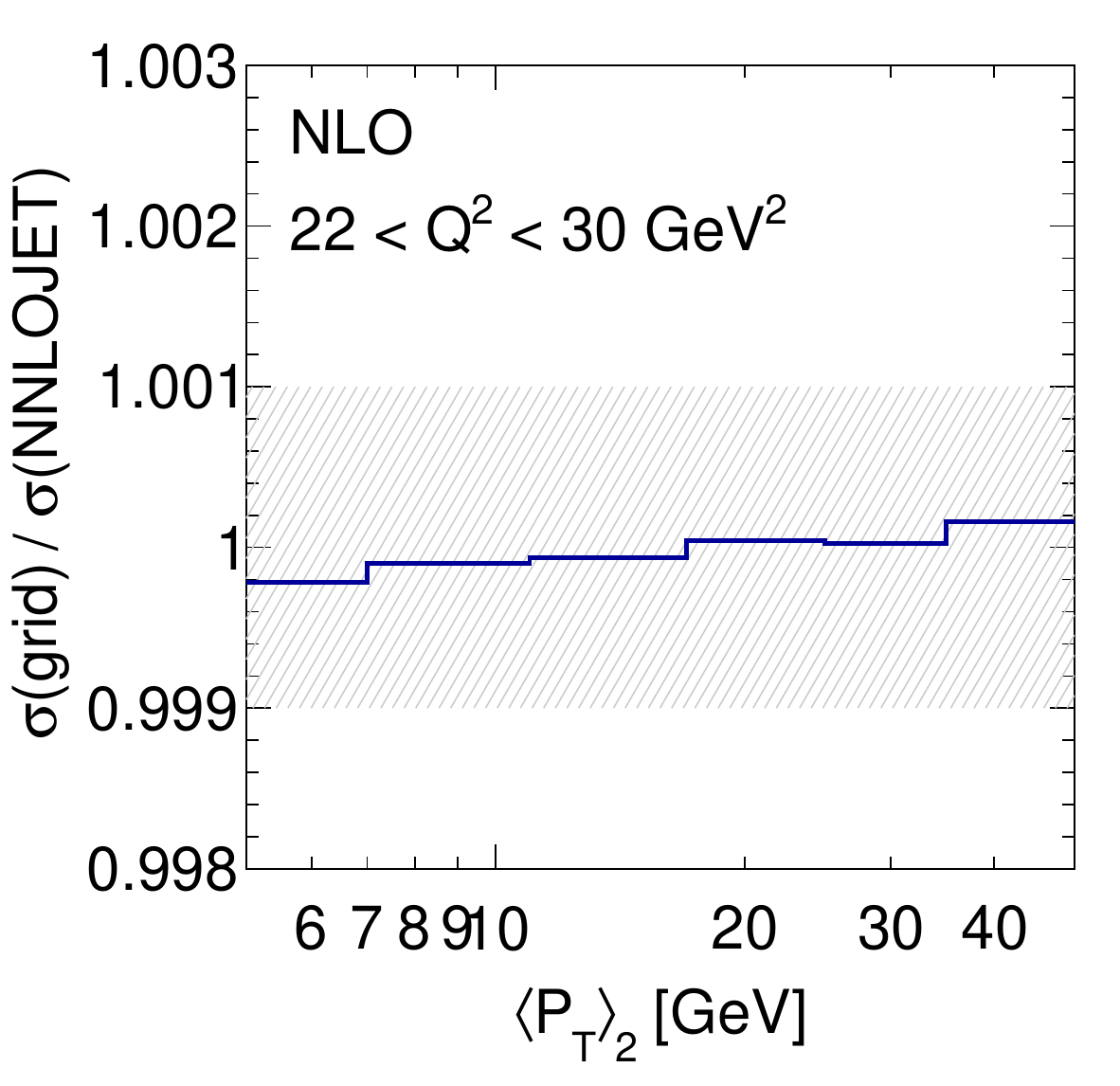}
  \includegraphics[width=0.31\textwidth]{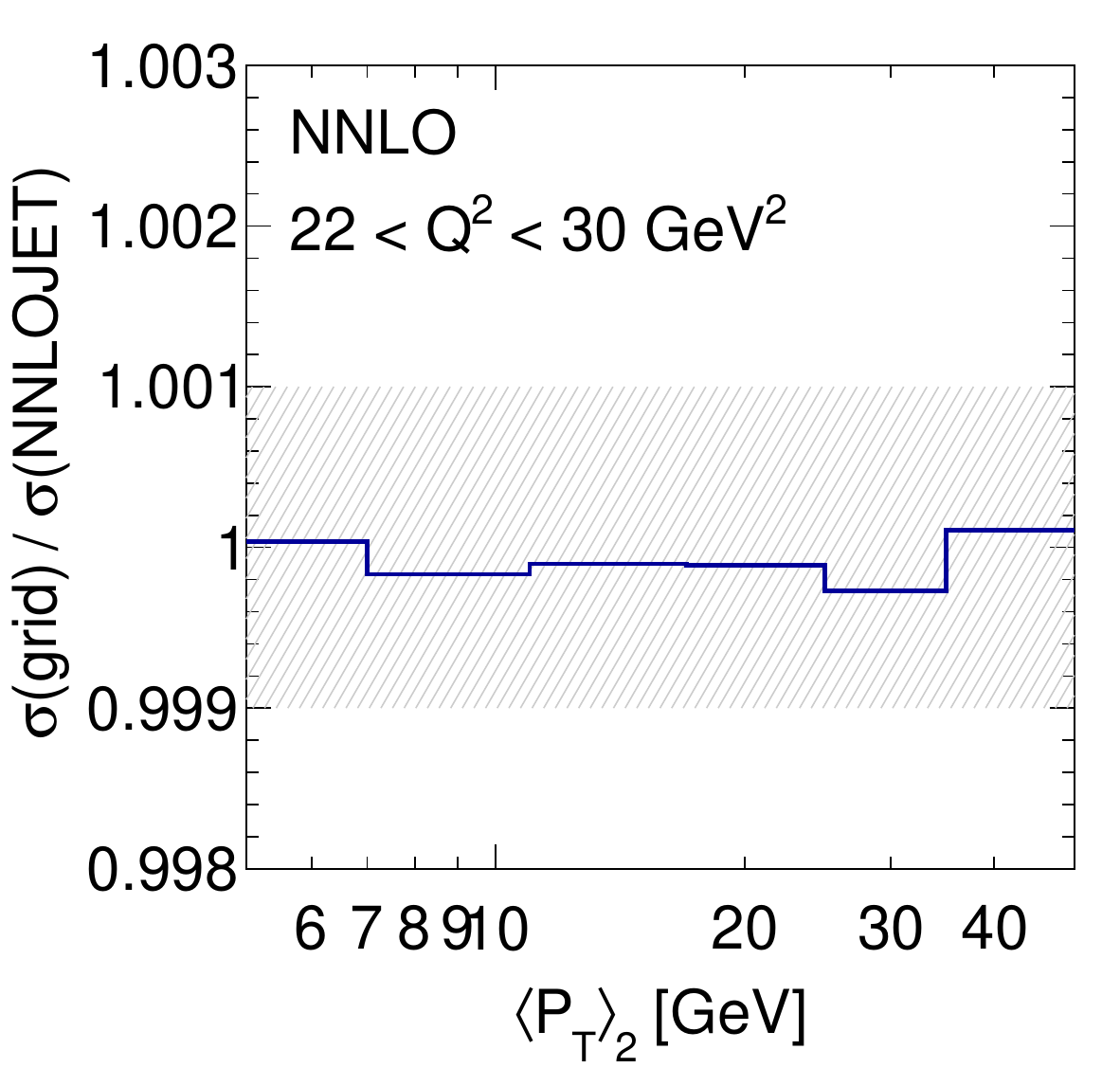}
  \\
  \includegraphics[width=0.31\textwidth]{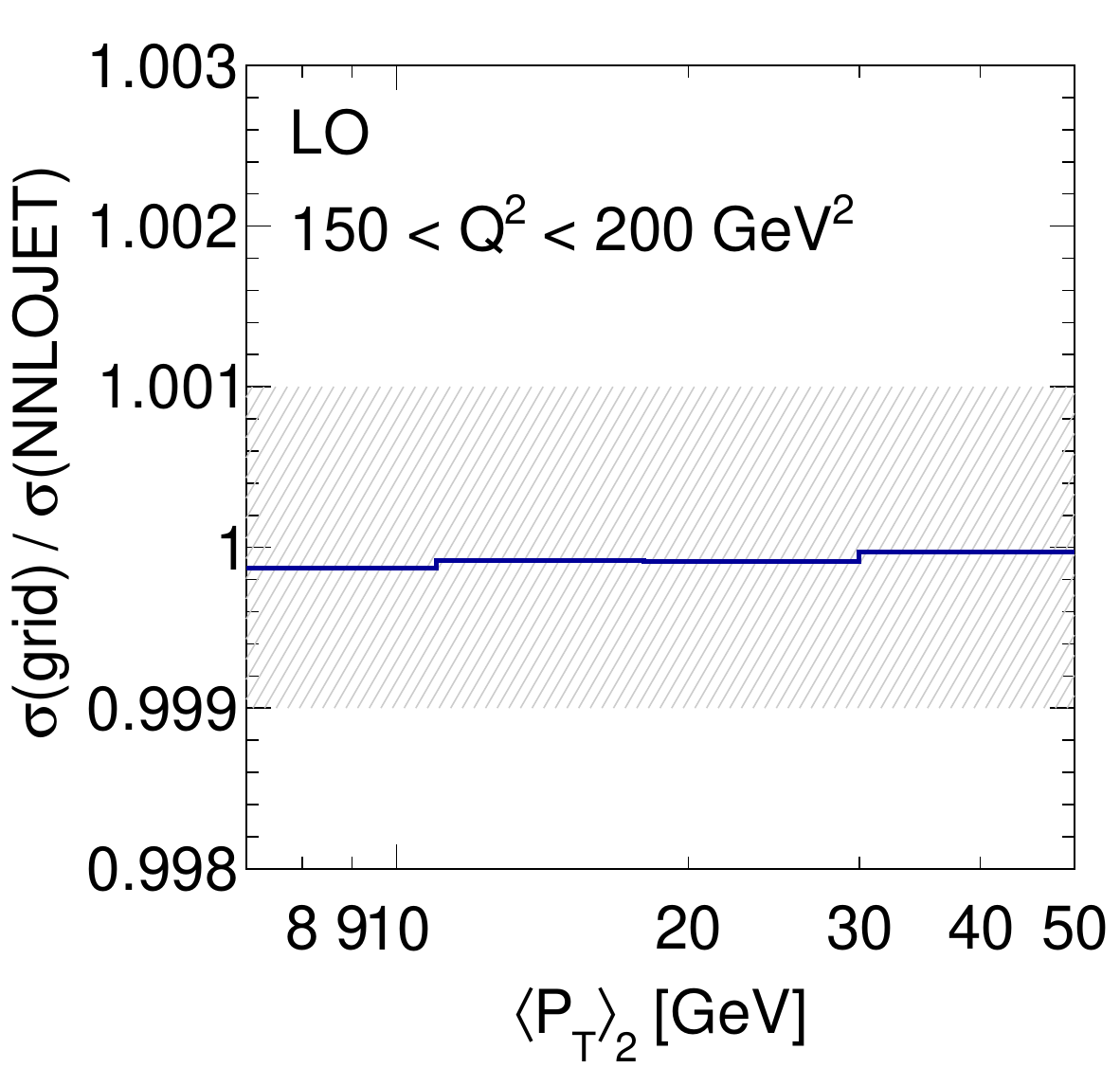}
  \includegraphics[width=0.31\textwidth]{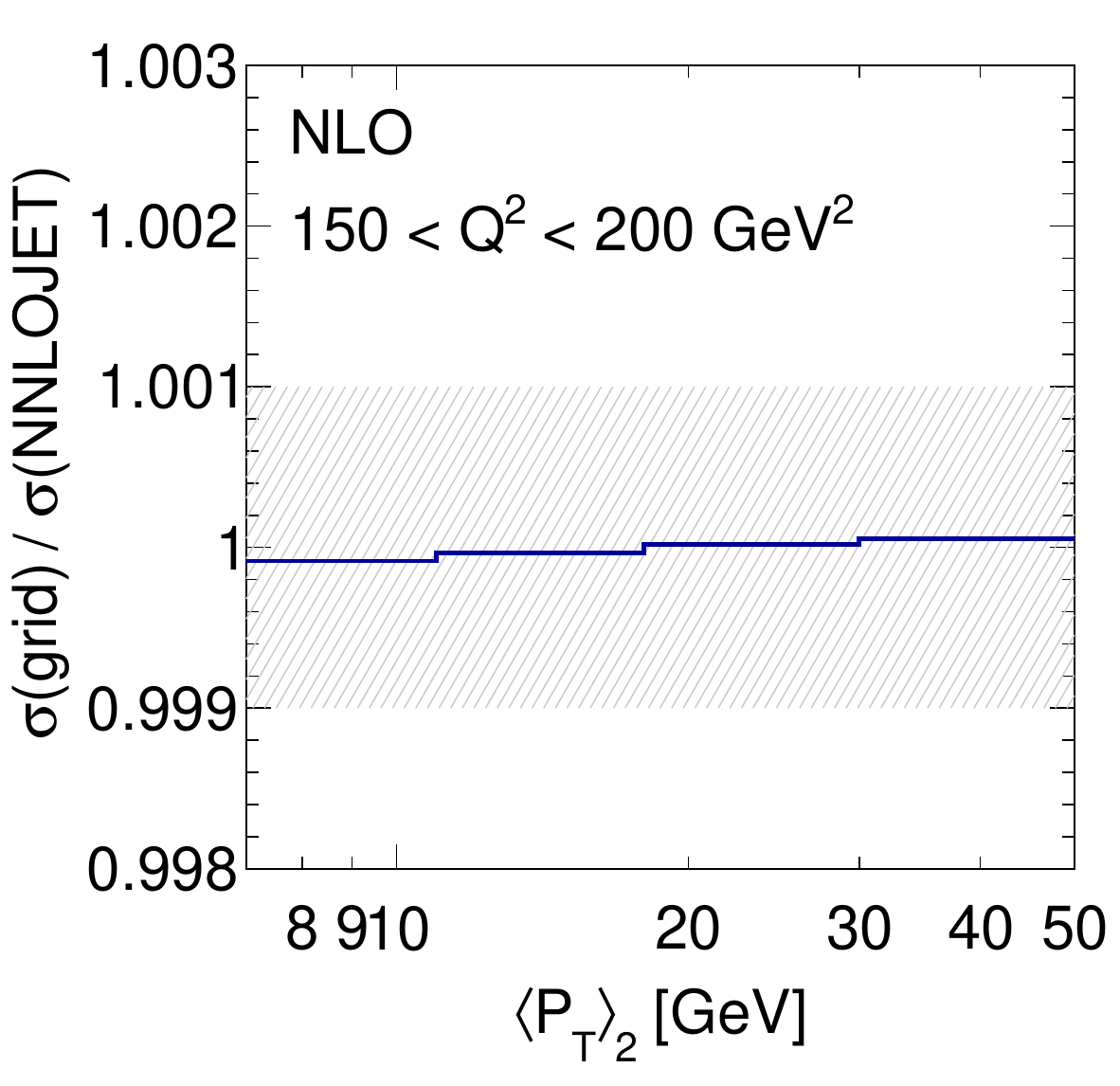}
  \includegraphics[width=0.31\textwidth]{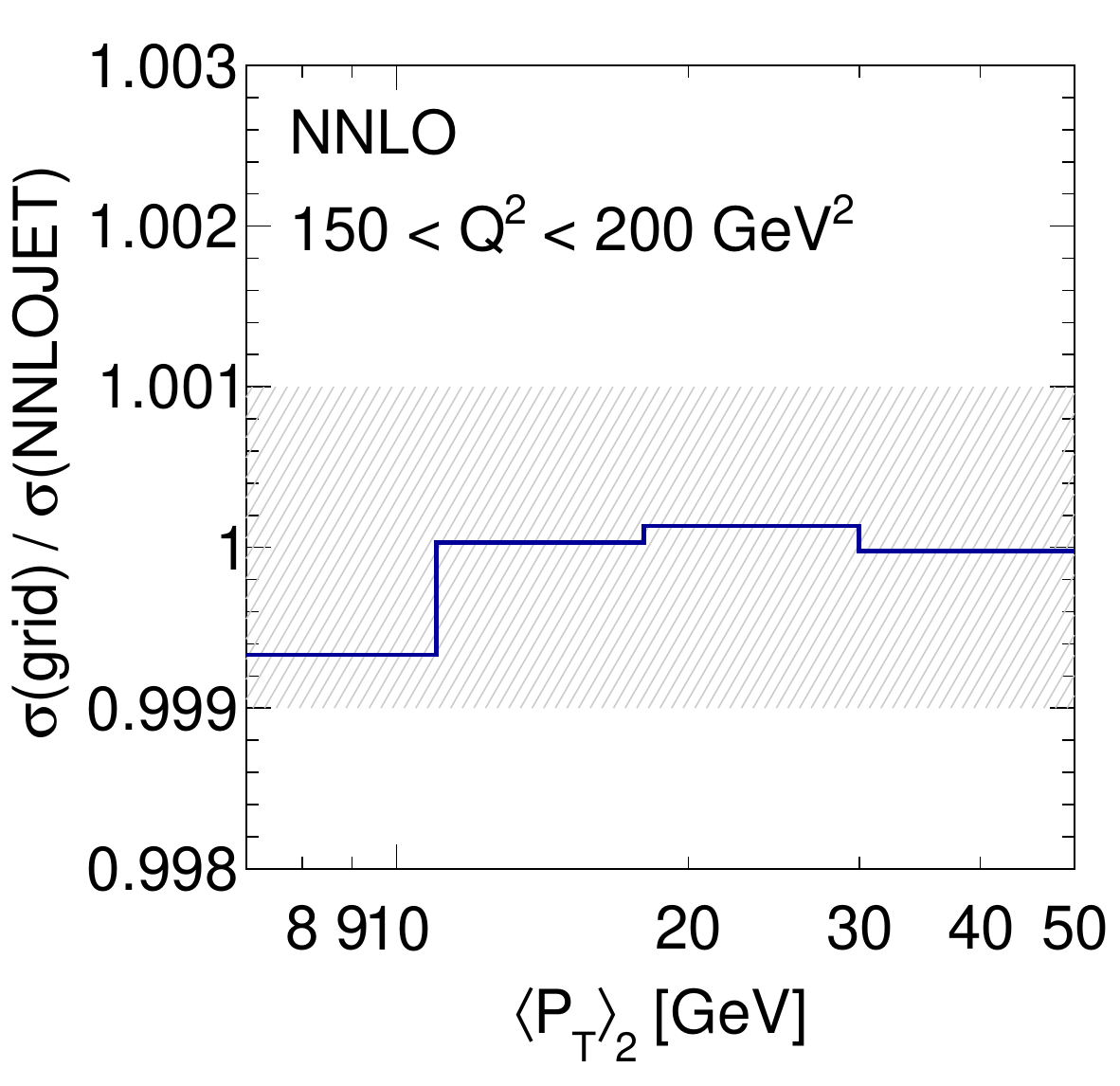}
  \caption{%
    Validation of the grid accuracy in di-jet production at
    low-\Qsq ($22<\Qsq<30\GeVsq$, top row) and high-\Qsq (\mbox{$150 <
      \Qsq < 200\GeVsq$}, bottom row).
    The shaded area indicates an agreement of 0.1\%.
    \label{fig:closure}
  }
\end{figure*}
}
\closefig

For a specific process, observable, and phase space selection, an initial
indication of the level of precision can be gained already using a single
job by comparing the interpolated result with
the reference calculation for the chosen PDF set for each bin in the observable.

Since identical events are filled both into the grid and into the
reference cross section, then any statistical fluctuations should be  reproduced
and thus a limited number of events is usually sufficient for this validation.
Subsequently, a
similar level of precision should be possible for each of the contributions
for the full calculation.
In future, this could be exploited to avoid the time consuming
access to the reference PDF during the full \nnlojet calculation itself
during the mass production of interpolation grids at a previously validated level of precision.

For the grids presented here, all events have been produced with reference
weights and the sufficiently accurate reproduction of the reference has been verified;
for each of the individual output grids from the many separate runs for
each contribution, for the combined grids from each contribution, and
for the final overall grid combination.  Figure~\ref{fig:closure}
compares the fast convolution with the reference from \nnlojet for
di-jet data at low $Q^2$ from H1~\cite{Andreev:2016tgi} and
demonstrates an agreement better than the per mille level for all
bins.

\begin{figure*}[ht]
  \centering
  \includegraphics[width=0.45\textwidth]{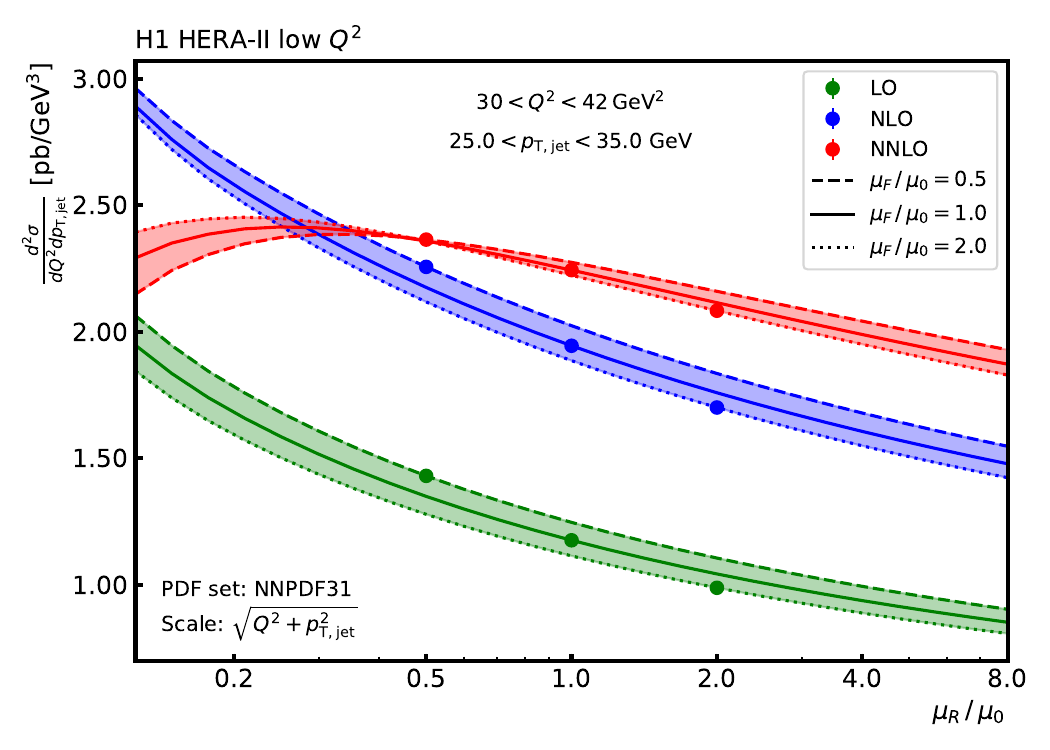}
  \includegraphics[width=0.45\textwidth]{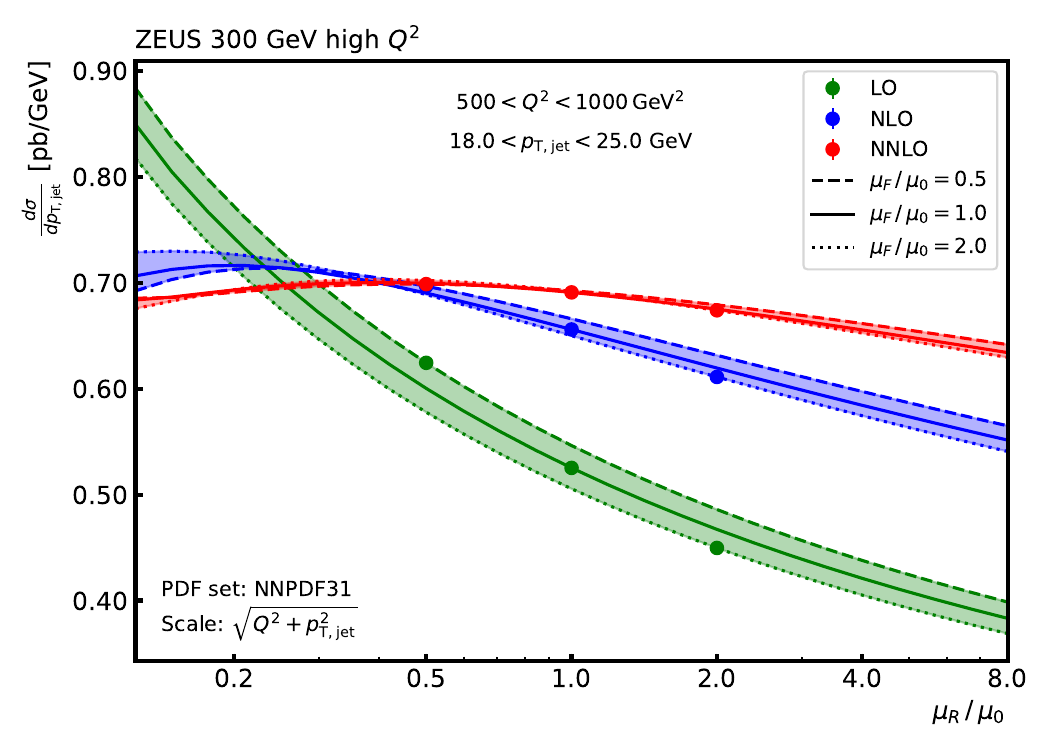}
  \caption{The scale dependence for a single bin in jet \pt with
    $25 < \ptjet < 35\GeV$ for a range
    $30 < \Qsq < 42\GeVsq$ from H1 (left) and in jet \pt with
    $18 < \ptjet < 25\GeV$ for a range
    $500 < \Qsq < 1000\GeVsq$ from ZEUS (right). The bands show the result of
    varying the factorisation scale $\muf$ by factors between 0.5 and
    2.0 with respect to the nominal scale. At each order three points
    indicate the result of symmetric variations of $\mur$ and $\muf$.}
  \label{fig:deps}
\end{figure*}

Additional cross checks can be performed, for example, comparing the
interpolated result of the final grid using an alternative PDF from
the reference cross section, with an independent reference
calculation for this same alternative PDF set. Here, of course, agreement
can only be confirmed within the statistical precision of the two
independent calculations. Moreover, it can be verified that the fast convolution with a
change in scale, $\mu$, is consistent with the full calculation performed at that scale.

In addition, the independent and completely different scale
variation techniques implemented in APPLgrid and
fastNLO are cross-checked against each other and are found
to agree. The resulting scale
dependence with a choice for the nominal scale of $\mu_0^2 = \Qsq+\ptjet^2$,
is illustrated in Figure~\ref{fig:deps} for two bins in
inclusive jet \pt; one from the H1 low \Qsq data and one for the ZEUS
high \Qsq data.

\begin{figure*}[ht]
  \centering
  \includegraphics[width=0.45\textwidth]{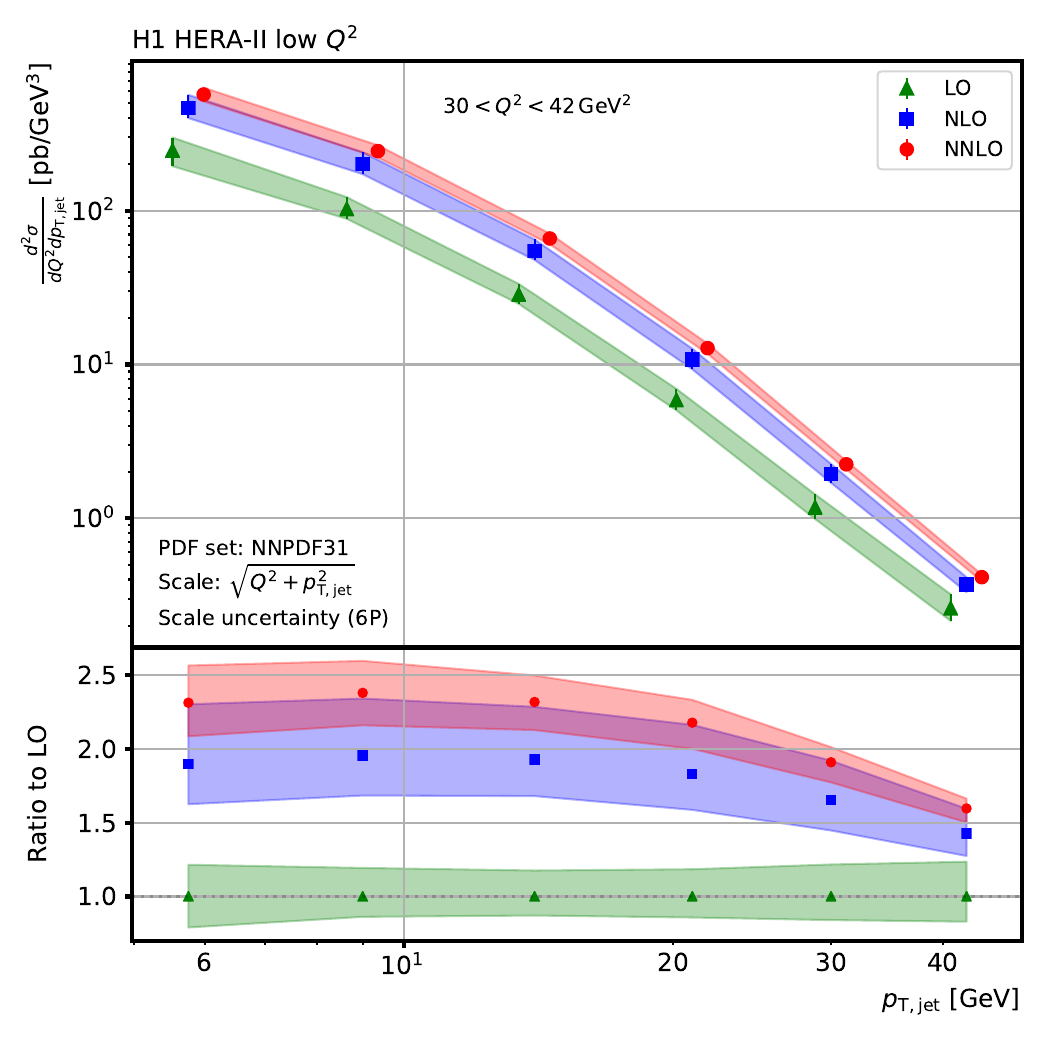}
  \includegraphics[width=0.45\textwidth]{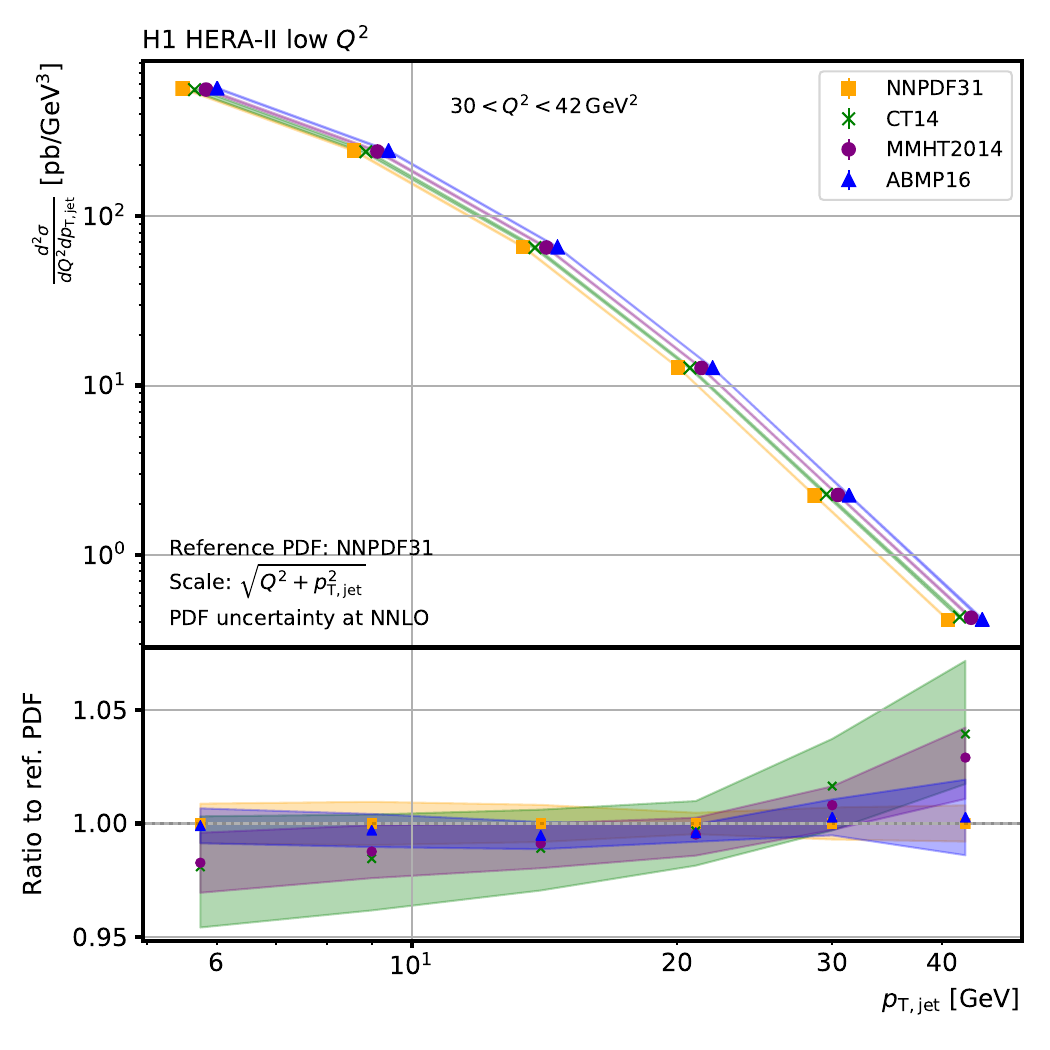}
  \includegraphics[width=0.45\textwidth]{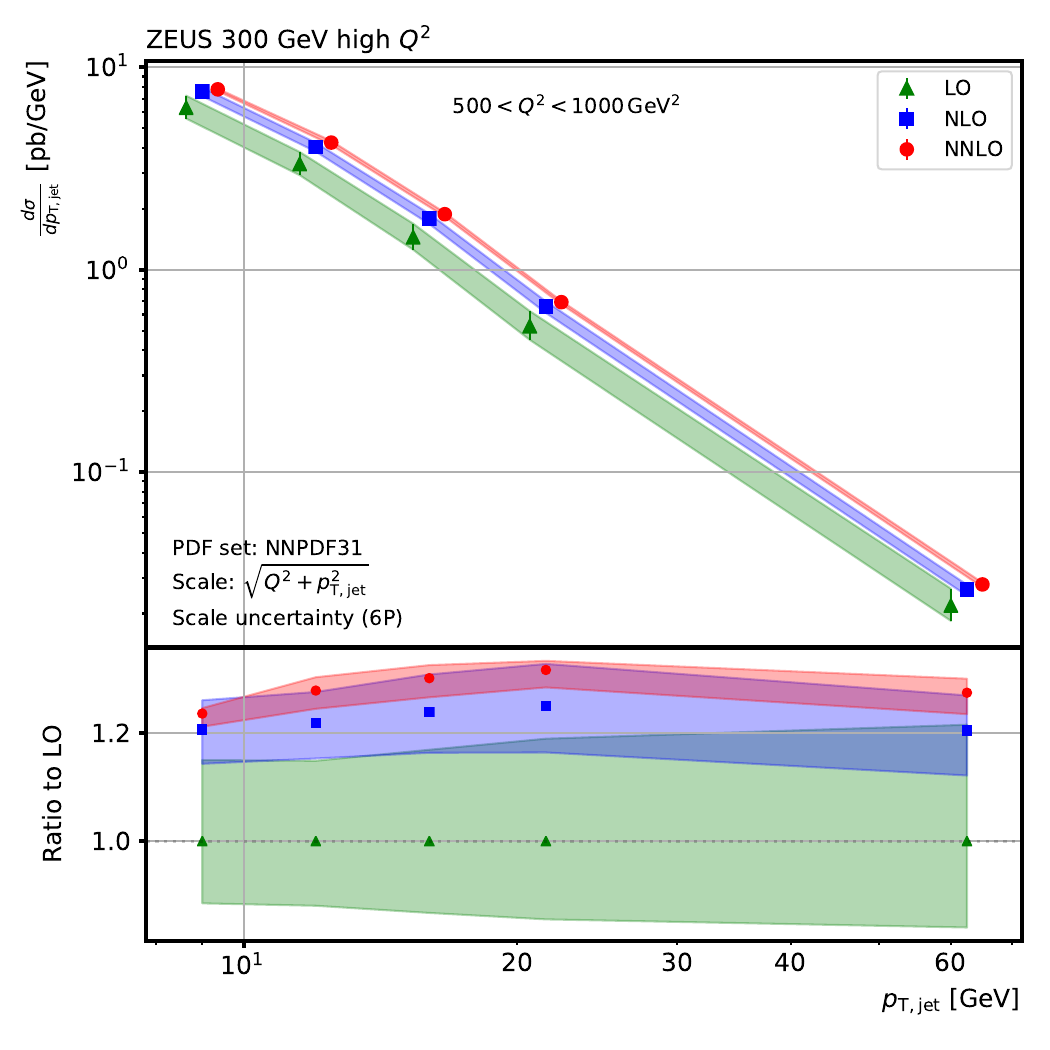}
  \includegraphics[width=0.45\textwidth]{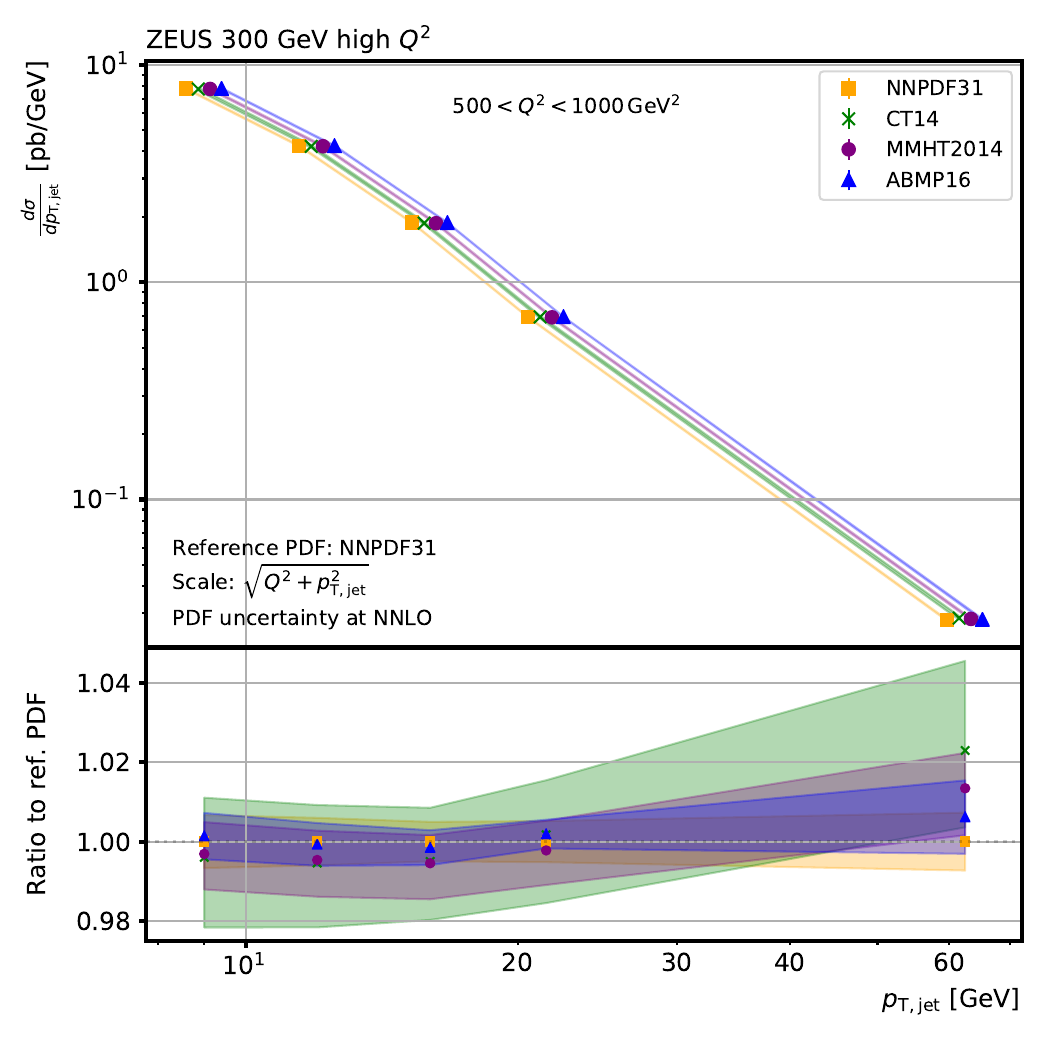}
  \caption{Inclusive jet cross section as a function of the jet \pt
    for two ranges in \Qsq: $30 < \Qsq < 42\GeVsq$ for H1 data (upper
    row), and $500 < \Qsq < 1000\GeVsq$ for ZEUS data (lower row). On
    the left the LO, NLO, and NNLO predictions are shown using the
    NNPDF31 PDF set including their ratio to the LO in the respective
    lower panels.  On the right the NNLO predictions are shown for the
    four PDF sets NNPDF31, CT14, MMHT2014, and ABMP16 including their
    ratio to the NNPDF31 PDF prediction in the respective lower
    panels.  The bands indicate the uncertainty derived from six
    variations of the $\mur$ and $\muf$ scale factors as
    described in the text (left), respectively the PDF uncertainty as
    prescribed in the respective publications. For better visibility
    the points in all upper panels are slightly shifted in \ptjet.}
  \label{fig:uncs}
\end{figure*}

A significant benefit of using such interpolation grids is that the
detailed uncertainties can be calculated without the need to rerun
the calculation. This is illustrated in
Figure~\ref{fig:uncs}, which shows the full seven point scale variation
and the PDF uncertainties
derived for the \ptjet dependent cross sections of the same
H1 and ZEUS measurements from before. The seven point scale uncertainty is a
conventional means of estimating the possible effect of uncalculated higher
orders. It is defined by the maximal upward and downward changes in
the cross section when varying the renormalisation and factorisation
scales by factors of two around the nominal scale in the following six
combinations of $(\mur/\mu_0, \muf/\mu_0)$: $(1/2, 1/2)$, $(2, 2)$,
$(1/2, 1)$, $(1, 1/2)$, $(2, 1)$, and $(1, 2)$. The PDF uncertainties
at the $1\,\sigma$ level are evaluated as prescribed for the respective
PDF sets\footnote{
  The full LHAPDF~\cite{Buckley:2014ana} names for each of the sets are: NNPDF31\_nnlo\_as\_0118,
  CT14nnlo, MMHT2014nnlo68cl, and ABMP16als118\_5\_nnlo respectively.
}
: NNPDF31~\cite{Ball:2017nwa},
CT14~\cite{Dulat:2015mca},
MMHT2014~\cite{Harland-Lang:2014zoa},
and ABMP16~\cite{Alekhin:2017kpj}.
In all plots PDFs at NNLO have been used with $\asmz=0.118$.%

%!TEX root = ../applfast.tex

% -----------------------------------------------------------------------
\section{Application: Determination of the strong coupling constant}
\label{qcd}
% -----------------------------------------------------------------------

As an application in using the DIS jet grids at NNLO,
an extraction of the
strong coupling constant, $\asmz$, is performed
using a fit of the NNLO QCD
predictions from \nnlojet to the HERA inclusive jet cross-section data.

% describe data sets
Seven sets of cross section measurements by the HERA
experiments are considered for the \asmz determination:
Five from H1 and two from ZEUS,
% At least one ZEUS table is not double-differential
each given by an inclusive jet cross
section measurement as a function of \ptjet and \Qsq.
The H1 results include measurements at $\sqrt{s}=300~\GeV$~\cite{Adloff:2000tq} and
$\sqrt{s}=320~\GeV$~\cite{Aktas:2007aa,Aaron:2010ac,Andreev:2014wwa,Andreev:2016tgi},
in the ranges $\Qsq\lesssim 120~\GeVsq$~\cite{Aaron:2010ac,Andreev:2016tgi}
and $\Qsq\gtrsim 120~\GeVsq$~\cite{Adloff:2000tq,Aktas:2007aa,Andreev:2014wwa},
where jets are measured within a kinematic
range between $4.5 <\ptjet<80~\GeV$.
For ZEUS, the data are similarly comprised of measurements at $\sqrt{s}=300~\GeV$~\cite{Chekanov:2002be} and
$\sqrt{s}=320~\GeV$~\cite{Chekanov:2006xr}, but in the range $\Qsq > 125~\GeVsq$ and with jets
having $\ptjet>8~\GeV$.
For all data sets jets are defined in the Breit frame of reference using the $k_T$ jet
algorithm with a jet-resolution parameter $R=1$.

% describe fit methodology
The methodology for the \asmz determination employs the same technique as
Refs.~\cite{Andreev:2017vxu} and~\cite{Britzger:2017maj}.
In brief, a goodness-of-fit quantifier between data and prediction
that depends on \asmz is defined in terms of a
\chisq function, which is based on normally-distributed relative
uncertainties and accounts for all experimental, hadronisation,
and PDF uncertainties, and any associated with the interpolation grids.
  The experimental uncertainties, and the hadronisation corrections and
  their uncertainties are provided together with the data
  by the H1 and ZEUS collaborations. The PDF uncertainties are calculated 
  using the prescriptions provided by the respective PDF fitting groups.
The \chisq function is then minimised
using Minuit~\cite{James:1975dr}.
The \asmz dependence in the predictions takes into account the contributions from both the hard coefficients
and the PDFs. The latter is
evaluated using the DGLAP evolution as implemented in the Apfel++
package~\cite{Bertone:2013vaa,Bertone:2017gds},
using the PDFs evaluated at a scale of $\mu_0=20~\GeV$.
A different choice for the value of $\mu_0$ is found to have
negligible impact on the results.
The uncertainties on the fit quantity are obtained by the HESSE
algorithm and validated by comparison with results obtained using the MINOS algorithm~\cite{James:1975dr}.
The uncertainties are separated into experimental (exp), hadronisation (had),
and PDF uncertainties (PDF) by repeating the fit excluding
uncertainty components.

Following Ref.~\cite{Andreev:2017vxu}, a representative value is assigned for the
renormalisation scale to each single data cross section measurement denoted by $\tilde{\mu}$.
This is determined from the lower and upper
bin boundaries in \Qsq and \ptjet (denoted with subscripts {\em dn} and
{\em up}) as
\begin{equation}
  \tilde{\mu}^2 = \sqrt{Q^{2}_{{\rm dn}} Q^{2}_{{\rm up}}} +
  p^{\rm jet}_{{\rm T,dn}}p^{\rm jet}_{{\rm T,up}}\,.
  \label{eq:mutilde}
\end{equation}
The calculation is performed using five
massless flavours, and as such, for the \as fit, the data are restricted to be above
twice the mass of the $b$-quark~\cite{Tanabashi:2018oca},
i.e.\ $\tilde{\mu}>2m_b$.

The nominal predictions are obtained using the NNPDF3.1 PDF
set~\cite{Ball:2017nwa}, which is used to further define the PDF and
PDF\as uncertainties. The PDFset uncertainties, on the other hand, are
determined by separately repeating the \as fit using predictions at
NNLO that are evaluated using the ABMP~\cite{Alekhin:2017kpj},
CT14~\cite{Dulat:2015mca}, HERAPDF2.0~\cite{Abramowicz:2015mha},
MMHT~\cite{Harland-Lang:2014zoa}, and NNPDF3.1 PDF sets.  The exact
definition of the PDF\as and PDFset uncertainties can be found in
Ref.~\cite{Britzger:2017maj}.

%/////////////////////
%// results: as(MZ) //
%/////////////////////

%/////////////////////
%// results: as(MZ) //
%/////////////////////

\begin{table*}[tbhp]
  \scriptsize
  \begin{center}
    \begin{tabular}{lllccc}
      \hline
            {\bf Data}         %
            & {\boldmath ~~$\tilmu_{\rm cut}$}
            & \multicolumn{1}{c}{\bf\boldmath \asmz with uncertainties}
            & \multicolumn{1}{c}{\bf th}
            & \multicolumn{1}{c}{\bf tot}
            & {\boldmath $\chisq/\ndf$}
            \\     %
            \hline
                {\bf H1 inclusive jets}${}^\dagger$ & & & &  &     \\
            $300\,\GeV$ high-\Qsq & $2m_b$
            & $ 0.1253\,(33)_{\rm exp}\,(23)_{\rm had}\,(5)_{\rm PDF}\,(3)_{\rm PDF\as}\,(5)_{\rm PDFset}\,(28)_{\rm scale}$  & $(37)_{\rm th}$  & $(49)_{\rm tot}$ & $3.7/15$  %   820-HQ-IJ
            \\
            HERA-I      low-\Qsq  & $2m_b$
            & $ 0.1113\,(18)_{\rm exp}\,~\,(8)_{\rm had}\,(5)_{\rm PDF}\,(5)_{\rm PDF\as}\,(7)_{\rm PDFset}\,(33)_{\rm scale}$  & $(36)_{\rm th}$  & $(40)_{\rm tot}$ & $14.6/22$  %   H-I-LQ-IJ
            \\
            HERA-I      high-\Qsq   & $2m_b$
            & $ 0.1163\,(26)_{\rm exp}\,~\,(9)_{\rm had}\,(6)_{\rm PDF}\,(4)_{\rm PDF\as}\,(3)_{\rm PDFset}\,(22)_{\rm scale}$  & $(25)_{\rm th}$  & $(36)_{\rm tot}$ & $13.2/23$  %   H-I-HQ-IJ
            \\
            HERA-II     low-\Qsq    & $2m_b$
            & $ 0.1212\,(16)_{\rm exp}\,(12)_{\rm had}\,(4)_{\rm PDF}\,(4)_{\rm PDF\as}\,(3)_{\rm PDFset}\,(38)_{\rm scale}$  & $(40)_{\rm th}$  & $(43)_{\rm tot}$ & $28.2/40$  %   HII-LQ-IJ
            \\
            HERA-II     high-\Qsq   & $2m_b$
            & $ 0.1156\,(20)_{\rm exp}\,(10)_{\rm had}\,(6)_{\rm PDF}\,(4)_{\rm PDF\as}\,(2)_{\rm PDFset}\,(24)_{\rm scale}$  & $(27)_{\rm th}$  & $(34)_{\rm tot}$ & $33.7/29$  %   HII-HQ-IJ
            \\
            \hline
%  Data-set   as(mz)      exp      had      PDF    PDFas   PDFset    scale   theory   total    chi^2  n_dof
% ZEUS-96-IJ   0.1240   0.0030   0.0003   0.0005   0.0001   0.0003   0.0038   0.0038  0.0049     27.5     29
%                         0.0038 (scale)  0.0017 (sclFac)  0.0027 (sclFac005)  0.0000 (scl-C)  0.0019 (scl-C/U)  0.0000 (PDFsum)  0.95   27.5   30
% ZEUS-98-IJ   0.1210   0.0029   0.0018   0.0005   0.0001   0.0004   -0.0013   0.0023  0.0037     17.9     29
%                         -0.0013 (scale)  0.0014 (sclFac)  -0.0015 (sclFac005)  0.0012 (scl-C)  0.0008 (scl-C/U)  0.0000 (PDFsum)  0.62   17.9   30
            {\bf ZEUS inclusive jets}  & & &       \\
            $300\,\GeV$ high-\Qsq & $2m_b$
  & $ 0.1240\,(30)_{\rm exp}\,~\,(3)_{\rm had}\,(5)_{\rm PDF}\,(1)_{\rm PDF\as}\,(3)_{\rm PDFset}\,(17)_{\rm scale}$  & $(18)_{\rm th}$  & $(35)_{\rm tot}$ & $26.9/29$  %  ZEUS-96-IJ
            \\
            HERA-I      high-\Qsq  & $2m_b$
  & $ 0.1211\,(29)_{\rm exp}\,(18)_{\rm had}\,(5)_{\rm PDF}\,(1)_{\rm PDF\as}\,(4)_{\rm PDFset}\,(14)_{\rm scale}$  & $(24)_{\rm th}$  & $(37)_{\rm tot}$ & $18.1/29$  %   ZEUS-98-IJ
            \\
              \hline
            {\bf H1 inclusive jets}${}^\dagger$  & & &       \\
            H1 inclusive jets  & $2m_b$
  & $ 0.1157\,(10)_{\rm exp}\,~\,(6)_{\rm had}\,(4)_{\rm PDF}\,(4)_{\rm PDF\as}\,(2)_{\rm PDFset}\,(34)_{\rm scale}$  & $(36)_{\rm th}$  & $(37)_{\rm tot}$ & $118.1/133$  %   IJ
            \\
            H1 inclusive jets & $28\,\GeV$
  & $ 0.1158\,(19)_{\rm exp}\,~\,(9)_{\rm had}\,(2)_{\rm PDF}\,(2)_{\rm PDF\as}\,(4)_{\rm PDFset}\,(21)_{\rm scale}$  & $(23)_{\rm th}$  & $(30)_{\rm tot}$ & $43.0/60$  %   IJ28
            \\
              \hline
%         ZS   0.1227   0.0021   0.0008   0.0005   0.0001   0.0004   0.0089   0.0090  0.0092     45.8     59
%                         0.0089 (scale)  0.0016 (sclFac)  0.0070 (sclFac005)  0.0012 (scl-C)  0.0010 (scl-C/U)  0.0000 (PDFsum)  0.78   45.8   60
%       ZS28   0.1209   0.0025   0.0005   0.0003   0.0002   0.0006   0.0018   0.0020  0.0032     34.1     43
%                         0.0018 (scale)  0.0015 (sclFac)  0.0005 (sclFac005)  0.0012 (scl-C)  0.0010 (scl-C/U)  0.0000 (PDFsum)  0.79   34.1   44
            {\bf ZEUS inclusive jets}  & & &       \\
            ZEUS inclusive jets  & $2m_b$
  & $ 0.1227\,(21)_{\rm exp}\,(9)_{\rm had}\,(6)_{\rm PDF}\,(1)_{\rm PDF\as}\,(4)_{\rm PDFset}\,(16)_{\rm scale}$  & $(19)_{\rm th}$  & $(28)_{\rm tot}$ & $45.5/59$  %   ZS
            \\
            ZEUS inclusive jets & $28\,\GeV$
  & $ 0.1208\,(25)_{\rm exp}\,(6)_{\rm had}\,(4)_{\rm PDF}\,(2)_{\rm PDF\as}\,(6)_{\rm PDFset}\,(15)_{\rm scale}$  & $(18)_{\rm th}$  & $(31)_{\rm tot}$ & $33.8/43$  %   ZS28
            \\
              \hline
%         HZ   0.1172   0.0009   0.0005   0.0004   0.0003   0.0003   0.0033   0.0034  0.0035    171.8    193
%                         0.0033 (scale)  0.0032 (sclFac)  0.0033 (sclFac005)  0.0030 (scl-C)  0.0022 (scl-C/U)  0.0000 (PDFsum)  0.89  171.8  194
%       HZ28   0.1179   0.0015   0.0007   0.0002   0.0002   0.0004   0.0019   0.0021  0.0026     80.3    104
%                         0.0019 (scale)  0.0019 (sclFac)  0.0019 (sclFac005)  0.0017 (scl-C)  0.0011 (scl-C/U)  0.0000 (PDFsum)  0.77   80.3  105
            {\bf HERA inclusive jets}  & & &       \\
            HERA inclusive jets  & $2m_b$
  & $ 0.1171\,~\,(9)_{\rm exp}\,(5)_{\rm had}\,(4)_{\rm PDF}\,(3)_{\rm PDF\as}\,(2)_{\rm PDFset}\,(33)_{\rm scale}$  & $(34)_{\rm th}$  & $(35)_{\rm tot}$ & $170.7/193$  %   HZ
            \\
            HERA inclusive jets & $28\,\GeV$
  & $ 0.1178\,(15)_{\rm exp}\,(7)_{\rm had}\,(2)_{\rm PDF}\,(2)_{\rm PDF\as}\,(4)_{\rm PDFset}\,(19)_{\rm scale}$  & $(21)_{\rm th}$  & $(26)_{\rm tot}$ & $79.2/104$  %   HZ28
            \\
            \hline
            \multicolumn{6}{l}{${}^\dagger$ previously fit in Ref.~\cite{Andreev:2017vxu}}
    \end{tabular}
    \caption{
      A summary of values of \asmz from fits to HERA inclusive jet cross
      section measurements using NNLO predictions.
      The uncertainties denote the experimental (exp), hadronisation
      (had), PDF, PDF\as, PDFset and scale uncertainties as described
      in the text.
      The rightmost three columns denote the quadratic sum of the
      theoretical uncertainties (th), the total (tot) uncertainties
      and the value of $\chisq/\ndf$ of the corresponding fit.
      }
    \label{tab:asresults}
    \end{center}
\end{table*}

\begin{figure}[tb]
    \begin{center}
      \begin{minipage}[t]{\textwidth}
      \includegraphics[width=0.48\textwidth]{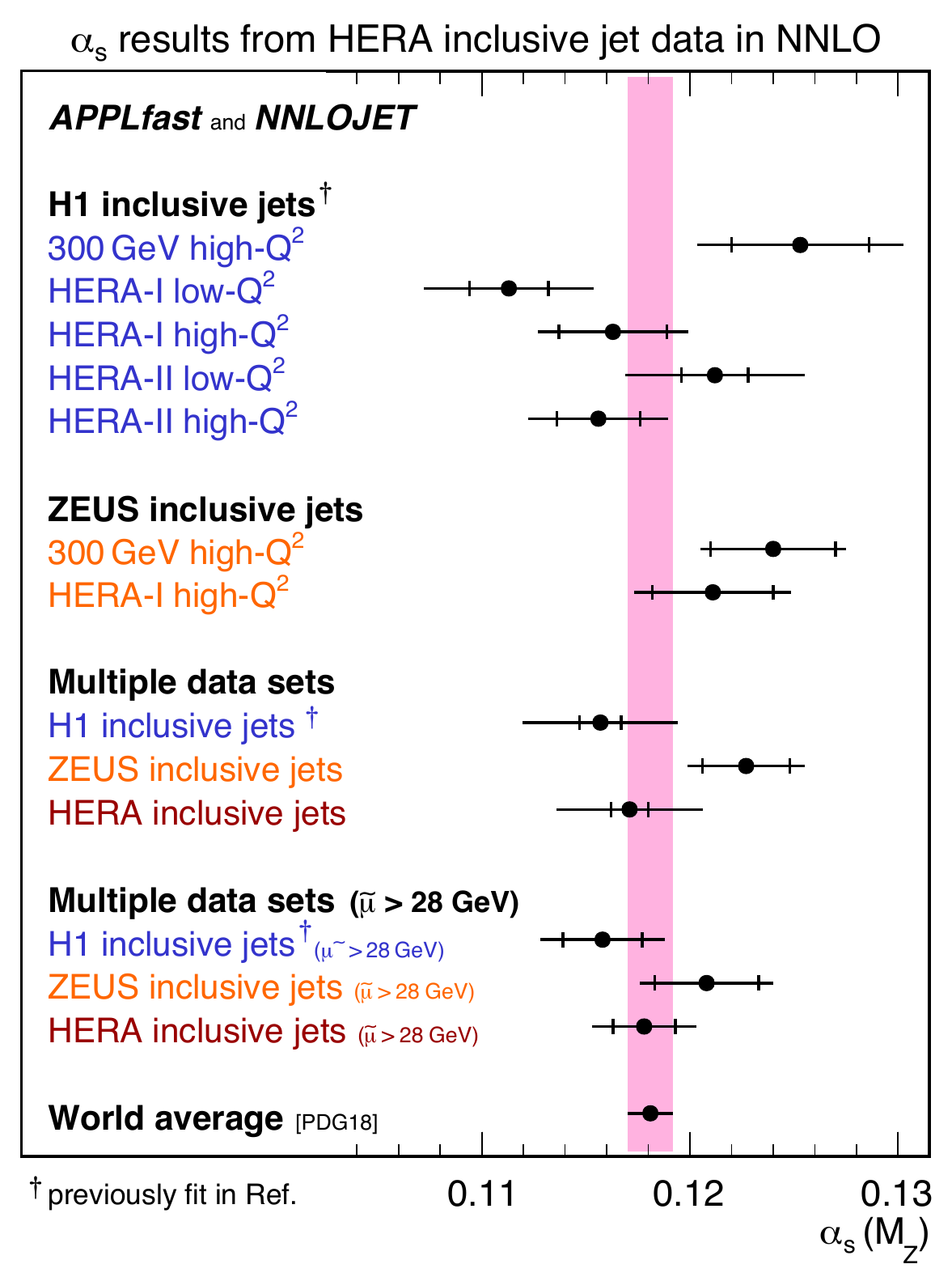}

      \vspace{-10.3mm}
      \font\sans = phvr  at 7.0pt  %scaled 1700
      \hspace{26.6mm}{\sans \cite{Andreev:2017vxu}}  \\
      \end{minipage}
      \caption{
        Summary of \asmz values in comparison with the world average value.
        The inner error bars indicate experimental uncertainties, and the full errors
        the total uncertainty, comprised of the experimental and theoretical uncertainties.
        The lower set of values represent fits to data restricted to $\tilde\mu>28\,\GeV$.
      }
      \label{fig:assummary}
    \end{center}
\end{figure}

% results, single data sets first
Results for the values of \asmz as obtained from the individual fits to the
inclusive jet cross section data are collected in
Table~\ref{tab:asresults}.
The entries for the H1 data sets correspond to
values previously reported in Ref.~\cite{Andreev:2017vxu} but some have been
updated using NNLO predictions with higher statistical precision.
New results are presented for the fits to the ZEUS inclusive jet cross
section data~\cite{Chekanov:2002be,Chekanov:2006xr} and
fits to all the H1 and ZEUS inclusive jet cross section data, which are the principle
results of this current study.
The \asmz values from the individual data sets are found to be
mutually compatible within their respective errors.
Figure~\ref{fig:assummary} summarises the values for
a visual comparison, and includes the world
average~\cite{PhysRevD.98.030001}, which is seen to
be consistent with the value extracted here.
All the H1 and ZEUS inclusive jet cross section data are found to be in good
agreement with the NNLO predictions, as indicated by the individual $\chisq/\ndf$ values in Table~\ref{tab:asresults}.
From the fit to all HERA inclusive jet data a value of
$\asmz=0.1171\,(9)_{\rm exp}\,(34)_{\rm th}$   %% bug-fixed value 2021!
is obtained, where {\em exp}  and {\em th} denote the
experimental and theoretical uncertainties, respectively, and where the latter is obtained
by combining individual theory uncertainties in quadrature.
A detailed description of the uncertainty
evaluation procedure can be found in Ref.~\cite{Andreev:2017vxu}.
The fit yields $\chisq/\ndf=170.7/193$, %% bug-fixed value 2021!
thus indicating an excellent
description of the data by the NNLO predictions.
Furthermore, an overall high degree of consistency for all of the HERA inclusive
jet cross section data is found.

% mu > 28GeV
The dominant uncertainty in the extraction of \as arises from
the renormalisation scale dependence of the NNLO predictions.
As such, the fits are repeated with a restricted
data selection requiring $\tilde{\mu}>28~\GeV$, chosen in order to
obtain a balance between the experimental uncertainty from the measurements
and the scale dependence from the theory predictions and so  reduce the total
uncertainty on the final extraction.
It was verified that the extracted $\as$ value and the associated uncertainty
are stable with respect to variations of $\tilde{\mu}$ around $28~\GeV$.
This fit represents the primary  result and the  value of \asmz is
determined to be
\begin{equation}
  \asmz = 0.1178\,(15)_\text{exp}\,(21)_\text{th}\,,  % bug-fixed value 2021
  \label{eq:as_value}
\end{equation}
with the uncertainty decomposition given in Table~\ref{tab:asresults}.
The value is found to be consistent with the world average
within uncertainties.
The obtained uncertainties are competitive with other determinations
from a single observable.

%/////////////////////////
%// results: as running //
%/////////////////////////
\begin{figure}[tb]
  \begin{center}
    \includegraphics[width=0.48\textwidth]{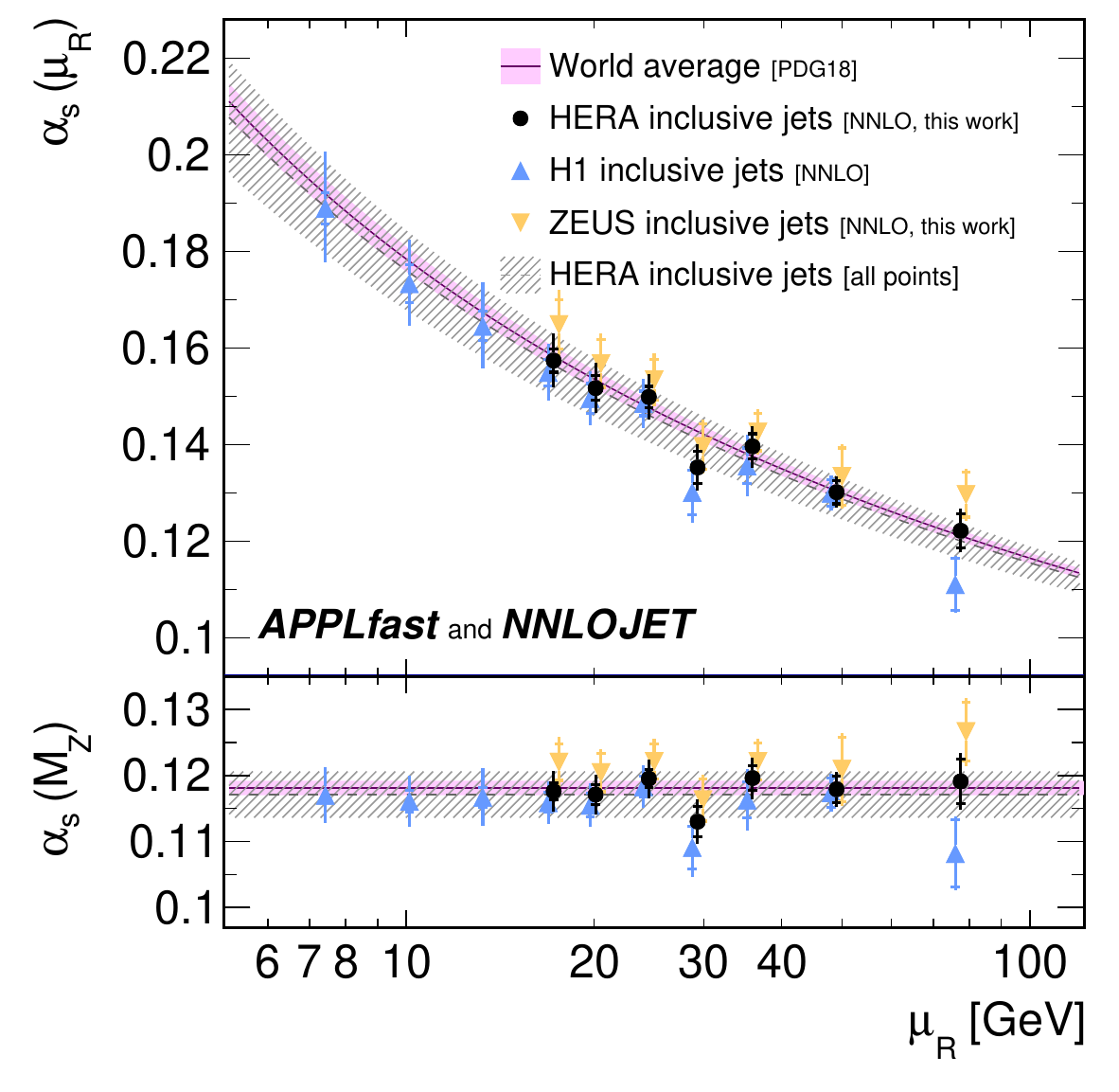}
    \caption{
      Results for \asmz (lower panel) and corresponding values for \asmur (upper panel)
      from fits to inclusive jet data points arranged in groups of similar \mur.
      The upper panel is obtained by applying the expectation from the QCD renormalisation group
      equation, as it also enters the NNLO predictions.
      The inner error bars indicate experimental uncertainties, and the full error bars
      the total uncertainty.
      The upper triangles show results from H1 data, which were previously fit in Ref.~\cite{Andreev:2017vxu}
      % which have first been
      % published in Ref.~\cite{Andreev:2017vxu}
      and are here partially updated with NNLO
      predictions with higher statistical accuracy.
      The lower triangles indicate the new results from ZEUS data.
      The full circles show the combined results from H1 and ZEUS data taken together and are labeled
      HERA inclusive jets.
      The shaded band indicates the world average value with its uncertainty, and
      the dashed line and hatched band indicate the result obtained from the fit to all inclusive jet data and its uncertainty.
    }
    \label{fig:asrunning}
  \end{center}
\end{figure}

%/////////////////////////
%// results: as values full stat grids
%/////////////////////////
%/////////////////////////
%// results: as values full stat grids
%/////////////////////////
\begin{table}[tbhp]
  %\footnotesize
  \scriptsize
  \begin{center}
    \begin{tabular}{cccc}
      \hline
            {\bf \mur}         %
%            {\bf $\tilde\mu$}         %
            & \multicolumn{1}{c}{H1}
            & \multicolumn{1}{c}{ZEUS}
            & \multicolumn{1}{c}{HERA}
            \\
            $[$GeV$]$
            & \multicolumn{1}{c}{\asmz}
            & \multicolumn{1}{c}{\asmz}
            & \multicolumn{1}{c}{\asmz}
            \\     %
            \hline
            7.4 % 7.47663. Now: 7.41889
 & $ 0.1170\,(13)\,(41)$     %   IJ  mu1
 & $ - $                     %   ZS  mu1
 & $ 0.1170\,(13)\,(41)$     %   HZ  mu1
            \\
            10.1 % 10.198. Now: 10.1193
 & $ 0.1161\,(17)\,(35)$     %   IJ  mu2
 & $ - $                     %   ZS  mu2
 & $ 0.1161\,(17)\,(35)$     %   HZ  mu2
            \\
            13.3 % 13.2665
 & $ 0.1167\,(15)\,(41)$     %   IJ  mu3
 & $ -$     %   ZS  mu3
 & $ 0.1167\,(15)\,(41)$     %   HZ  mu3
            \\
            17.2 % 17.2337
 & $ 0.1161\,(15)\,(28)$     %   IJ  mu4
 & $ 0.1220\,(28)\,(26)$     %   ZS  mu4
 & $ 0.1176\,(13)\,(28)$     %   HZ  mu4
            \\
            20.1 % 20.1246
 & $ 0.1158\,(18)\,(28)$     %   IJ  mu5
 & $ 0.1204\,(29)\,(22)$    %   ZS  mu5
 & $ 0.1171\,(15)\,(26)$     %   HZ  mu5
            \\
            24.5 % 24.4949
 & $ 0.1184\,(16)\,(27)$     %   IJ  mu6
 & $ 0.1221\,(27)\,(22)$    %   ZS  mu6
 & $ 0.1195\,(14)\,(26)$     %   HZ  mu6
            \\
            29.3 % 29.3258
 & $ 0.1091\,(32)\,(31)$     %   IJ  mu7
 & $ 0.1163\,(32)\,(20)$    %   ZS  mu7
 & $ 0.1130\,(23)\,(24)$     %   HZ  mu7
            \\
            36.0 % 35.9166
 & $ 0.1164\,(27)\,(38)$     %   IJ  mu8
 & $ 0.1221\,(28)\,(19)$    %   ZS  mu8
 & $ 0.1196\,(19)\,(26)$     %   HZ  mu8
            \\
            49.0 % 48.9898
 & $ 0.1174\,(22)\,(17)$     %   IJ  mu9
 & $ 0.1208\,(48)\,(27)$     %   ZS  mu9
 & $ 0.1179\,(20)\,(18)$     %   HZ  mu9
            \\
            77.5 % 77.4597
 & $ 0.1082\,(51)\,(22)$     %   IJ  mu10
 & $ 0.1266\,(44)\,(26)$    %   ZS  mu10
 & $ 0.1191\,(34)\,(26)$     %   HZ  mu10
            \\
            \hline
    \end{tabular}
    \caption{
      Values of the strong coupling constant at the $\PZ$-boson
      mass, \asmz, obtained from fits to groups of data with
      comparable values of $\mur$.
      The first (second) uncertainty of each point corresponds to the
      experimental (theory) uncertainties. The theory uncertainties include
      PDF related uncertainties and the dominating scale uncertainty.
    }
    \label{tab:asrunning}
    \end{center}
\end{table}

%% Running
The running of \asmur can be inferred from separate fits to groups of
data points that share a similar value of the renormalisation scale, as
estimated by $\tilde{\mu}$ in Eq.~\eqref{eq:mutilde}.
To this end, the \asmz values are determined for each $\tilde{\mu}$ collection individually,
and  are summarised in
Table~\ref{tab:asrunning} and shown in the bottom panel of
Fig.~\ref{fig:asrunning}.
All values are mutually compatible and in good agreement with the
world average, and no significant dependence on \mur is observed.
The corresponding values for $\asmur$, as determined using
the QCD renormalisation group equation, are
displayed in the top panel of Fig.~\ref{fig:asrunning}, illustrating the running of the strong
coupling.
The dashed line corresponds to the prediction for the \mur dependence using the \as value of
Eq.~\eqref{eq:as_value}.
The predicted running is in excellent agreement with the individual \asmur determinations,
further reflecting the internal consistency of the study.

% Conclusion
To conclude this study it is worth commenting on the robustness of the procedure.
On the theory side, the inclusive jet cross section represents an observable
that is well defined in perturbative QCD and only moderately affected by non-perturbative effects and experimentally,
this study rests on a solid basis, making use of
measurements from two different experiments based on three separate
data taking periods, which cover two different centre-of-mass energies
and two kinematic regions in \Qsq.
As a result, although only a single observable is used in the determination of \as,
a highly competitive experimental and theoretical precision is achieved.

%!TEX root = ../applfast.tex

% -----------------------------------------------------------------------
\section{Conclusions and Outlook}
\label{conclusions}
% -----------------------------------------------------------------------

NNLO calculations in perturbative QCD are rapidly becoming the new standard for many important scattering processes.
These calculations are critical in reducing theory uncertainties and often improve the description of the increasingly precise
data,
sometimes even resolving prior tensions.
However, the computational resources required for such calculations prohibit their use in applications that
require a frequent re-evaluation using different input conditions, e.g.\ fitting procedures for PDFs and Standard Model parameters.

Fast interpolations grid techniques circumvent these limitations by allowing for the {\em a posteriori} interchange
of PDFs, values of the strong coupling \as, and scales in the prediction at essentially no cost.
In this article the APPLfast project is discussed, which provides a generic interface for the APPLgrid and
fastNLO grid libraries to produce interpolation tables where the hard coefficient functions are computed by the \nnlojet program.
Details on the extension of the techniques to NNLO accuracy and their implementation for DIS are discussed, together with
the public release of NNLO grid tables for jet cross-section measurements at HERA~\cite{DIS:grids}.

As an application of the grids, an extraction of the strong coupling constant \as has been performed, based on jet data at HERA,
closely following the methodology in Refs.~\cite{Britzger:2017maj,Andreev:2017vxu}.
In contrast to Ref.~\cite{Andreev:2017vxu}, where the \as determination considered both inclusive and di-jet cross section
data from H1 alone,
this current analysis includes data from both the H1 and ZEUS experiments, but \as is fitted solely using the single jet inclusive data.
The usage of a single observable facilitates the simultaneous
determination of \asmz from two experiments, as the observable is
defined identically between both experiments and thus reduces
ambiguities in the treatment of theory uncertainties.
This work represents one of the first determinations of the strong coupling
constant to include both H1 and ZEUS DIS jet data at NNLO accuracy, 
where such a determination is only possible using the foundational work 
presented in this paper. The determination of \asmz from H1 and ZEUS data taken together provides
a best-fit value of $\asmz = 0.1178\,(15)_\text{exp}\,(21)_\text{th}$. %% bugfixed value 2021

Although the discussion in the present work was limited to the DIS
process, the implementation in both APPLfast and \nnlojet is fully
generic and thus generalisable to hadron-hadron collider processes.
This means that all NNLO calculations available from within \nnlojet,
such as di-jet production and $V+\jet$ production in proton-proton
scattering, are interfaced to grid-filling tools in a rather
straightforward manner.  This generalisation will be presented in a
future publication.

%!TEX root = ../applfast.tex

% -----------------------------------------------------------------------
\section*{Acknowledgements}
% -----------------------------------------------------------------------
This research was supported in part by the UK Science and Technology Facilities Council, by the Swiss National Science Foundation (SNF) under contracts 200020-175595 and 200021-172478, by the Research Executive Agency (REA) of the European Union through the ERC Advanced Grant MC@NNLO (340983) and the Funda\c{c}\~{a}o para a Ci\^{e}ncia e Tecnologia (FCT-Portugal), under projects UID/FIS/00777/2019, CERN/FIS-PAR/0022/2017. 
CG and MS were supported by the IPPP Associateship program for this project.
JP gratefully acknowledges the hospitality and financial support of the CERN theory group where work on this paper was conducted.

% \clearpage
\bibliography{applfast_red}

\end{document}